\documentclass[12pt]{article}
\usepackage{epsfig}
\normalbaselines
\newcommand{\be}{\begin{eqnarray}}
\newcommand{\ee}{\end{eqnarray}}

\def\ll#1{\left#1}
\def\r#1{\right#1}
\def\fr{\frac{1}{2}}

\def\mref#1{(\ref{#1})}

\def\p{\partial}

\def\bd{\begin{displaymath}}
\def\ed{\end{displaymath}}

\def\ba#1{\begin{array}{#1}}
\def\ea{\end{array}}
\def\nn{\nonumber}

\begin{document}

\pagestyle{empty}

\begin{center}

{\LARGE\bf The high energy semiclassical asymptotics of loci of roots of fundamental solutions for polynomial potentials\\[.5cm]}

\vskip 60pt

{\large {\bf Stefan Giller}}

\vskip 18pt

Jan Dlugosz Academy in Czestochowa\\
Institute of Physics\\ul. Armii Krajowej 13/15, 42-200 Czestochowa, Poland\\
e-mail: sgiller@uni.lodz.pl
\end{center}

\vspace{3 cm}

\begin{abstract}In the case of polynomial potentials all solutions to 1-D Schr\"odinger equation are entire functions
totally determined by loci of their roots and their behaviour at infinity. In this paper a description of the first of the
two properties is given for fundamental solutions for the high complex energy limit when the energy is quantized or not.
In particular due to the fact that the limit considered is semiclassical it is shown that loci
of roots of fundamental solutions are collected of selected Stokes lines (called exceptional) specific for the solution considered and are
distributed along these lines in a
specific way. A stable asymptotic limit of loci of zeros of fundamental solutions on their exceptional Stokes lines has island
forms and there are
infintely many of such roots islands on exceptional Stokes lines escaping to infinity and a finite number of them on exceptional Stokes
lines which connect pairs of
turning points. The results obtained for asymptotic roots distributions of fundamental solutions in the semiclassical high
(complex) energy limit are of a general nature for polynomial potentials.

\vskip 36pt

PACS number(s): 03.65.-W , 03.65.Sq , 02.30.Lt , 02.30.Mv

MSC classes:  34B05; 34L20; 34M40; 34M60


\end{abstract}

\newpage

\pagestyle{plain}

\setcounter{page}{1}

\section{Introduction}

\hskip+1.5em As it is already well known \cite{1,2} the fundamental solutions (FS) \cite{3} have appeared
to play main role in one dimensional quantum mechanics (or in a multi dimensional one allowing a reduction to the one
dimension) with analytic potentials, i.e. polynomial, meromorphic etc. In particular they allow us to solve all basic problems
typical for the field - eigenvalue problems, scattering problems, problems of decaying and JWKB and adiabatic limits
\cite{4}.
They are exceptional also
among all the solutions to the corresponding Schr\"odinger equations because of their property of being Borel summable
for the polynomial or meromorphic potentials \cite{5}. The latter property allows us to recover these solutions from their JWKB
series, both exactly and approximately if the semiclassical series is abbraviated in the latter case. Finally they allow us
to use in effective calculations their approximations by JWKB formulae having these approximations under full control.

Not surprising therefore that the fundamental solutions are also very important for mathematicians, used by them under the
name of subdominant solutions, to
study properties of solutions to Schr\"odinger equations (SE) for polynomial and meromorphic potentials \cite{6}. Among many
problems to be under considerations is the one of loci of zeros of solutions to SE as a function of coefficients of polynomial
or meromorphic (rational) potentials (see for example \cite{7}). This problem is also important from the physical point of view to mention
the known relation between
a number of real zeros of the quantized solution to SE and a number of an energy level corresponding to the solution. In
a recent paper of Bender {\it et al} \cite{11} a method of looking for zeros of eigenfunctions of eigenvalue problems with
non-hermitian potentials was suggested as a tool for checking a possible completeness of the full set of such eigenfunctions.

In another recent paper of Eremenko {\it et al} \cite{8} the high energy limit has been
considered to study this problem for a quantized energy in polynomial potentials. The
authors have shown that in
this limit the problem simplifies greatly so that it can be standardized and the loci of
zeros of
the corresponding quantized FS's (i.e. these which are a solution to an eigenvalue problem)
are exactly
on Stokes lines, called exceptional by the authors, of a Stokes graph (or global Stokes
lines - a name used by mathematiciens) corresponding to a considered problem.

Another semiclassical limit has been considered by Hezari \cite{9} who investigated the
problem of complex zeros of eigenfunctions of SE with real polynomial potentials of even degree
in the limit $\hbar\to 0$, where $\hbar$ is
the Planck constant, while the energy parameter $E$ was kept fixed.

However, Hezari's eigenfunctions are as such for
a problem of the quantized Planck constant (being by this a little bit unphysical). In fact the two cases, i.e. the
energy quantized while the Planck constant is kept fixed and the energy kept fixed but the Planck constant is quantized
have two
different semiclassical limits for the quantized parameters, i.e. the high energy limit and the small $\hbar$-limit lead
to different behaviour of the corresponding Stokes graphs and to different sets of eigenfunctions. Nevertheless since
both these limits are of the same semiclassical nature at least mathematically it is therefore not surprising that
Hezari's results of complex zeros eigenfunctions problem are similar to those of Eremenko {\it et al}. We have discussed
and generalized Hesari's results in a separate paper \cite{10}

In this paper we would like both to generalize the results of  Eremenko {\it et al} \cite{8} and to make the corresponding
theorems more precise in the following aspects:
\begin{enumerate}
\item to show that they are valid for not quantized FS's in the considered limit so that
quantized cases can be seen as particular results of these general ones
\item to show how the quantization procedure modifies unquantized zeros distributions of FS's
\item to show that the exceptional Stokes lines of Eremenko {\it et al} \cite{8} can be identified with the boundaries of
the canonical domains corresponding to FS's
\item to find the limit distributions
of zeros of FS's along exceptional SL's using the explicite form of FS's and their high energy semiclassical limit.
\end{enumerate}

In fact our analysis provides a full description of the root distribution problem of FS's to SE with polynomial
potentials in the high (complex) energy limit.

The paper is organized as follows.

In the next section the high energy limit of SE for polynomial potentials is considered and it standard form is
established.

In section 3 Stokes graphs (SG) for the standardized high energy polynomial potential are considered and their possible
standard forms are established in this limit.

In the next section we define the full set of fundamental solutions for the standard high energy limit potential
$(-i\alpha z)^n$ and formulate the basic lemma on possible posisions of roots of the FS's.

In section 5 two theorems are formulated establishing the precise positions of roots of FS's in the high energy limit.

In section 6 the results of the previous section are extrapolated to an arbitrary polynomial potential rescaled
correspondingly to the limit considered.

In section 7 the quantized cases of FS's and an influence of the quantization on distributions of zeros of FS's are
investigated.

We conclude and summarize the results in section 8.

\section{High energy limit of Stokes graphs for polynomial potentials}

\hskip+2em Consider the stationary 1-D Schr\"odinger equation with a polynomial potential
$P_n(z)=a_nz^n+...+a_1z,\;a_n\neq 0,\;n>1$, (a free term is assumed to be absorbed by the energy parameter $E$). We have:
\be
\phi''(z)-\frac{2m}{\hbar^2}(P_n(z)-E)\phi(z)=0
\label{1}
\ee

To standardize our problem we make the substitution $z\to -i\alpha\ll(\frac{E}{a_n}\r)^{\frac{1}{n}}z,\;\alpha=1$ for $n$ odd and
$\alpha=1,e^{-i\frac{\pi}{n}}$ for $n$ even so that we get for
$\psi_\alpha(z)\equiv\phi\ll(-i\alpha\ll(\frac{E}{a_n}\r)^{\frac{1}{n}}z\r)$
\be
\psi_\alpha''(z)-\lambda^2 W_n(z,\lambda)\psi_\alpha(z)=0
\label{2}
\ee
where
\be
W_n(z,\lambda)\equiv (-i\alpha z)^n-1+
\sum_{k=1}^{n-1}b_{n-k}(-i\alpha)^\frac{2k}{n+2}\lambda^{-\frac{2k}{n+2}}(-i\alpha z)^{n-k}\nn\\
\lambda^2=-\alpha^2\frac{2m}{\hbar^2}a_n^{-\frac{2}{n}}E^\frac{n+2}{n},\;\;\;\;
b_{n-k}=\frac{a_{n-k}}{a_n^{\frac{n-k+2}{2(n+2)}}}\ll(\frac{2m}{\hbar^2}\r)^\frac{k}{n+2},\;\;\;k=1,...,n-1
\label{3}
\ee
where the factor $-i\alpha$ has been introduced for a convenience what will be self-explaining later.

Let us remind that for the case considered the Stokes graph (SG) is created as a set of all lines (Stokes lines (SL))
emerging from each root $z_i(\lambda),\;i=1,...,n$, of $W_n(z,\lambda)$ and satisfying one of the following equations
\be
\Re\ll(\lambda \int_{z_{i}(\lambda)}^{z} \sqrt{W_n(\xi,\lambda)}d\xi\r) = 0,\;\;i=1,...,n
\label{4}
\end{eqnarray}

The roots $z_i(\lambda),\;i=1,...,n$ will be called also turning points.

From \mref{3} it is obvious that for $|E|\to\infty$, i.e. $|\lambda|\to\infty$, the limit Stokes graph corresponding to
the rescaled problem is determind by the roots (turning points) of $W_{n,\alpha}(z)\equiv(-i\alpha z)^n-1$ and the phase of
$\lambda$.
It is also clear that for the considered limit the roots $z_i(\lambda)$ are all simple and their loci are close
to the ones of $W_{n,\alpha}(z)$ so that the SG corresponding to $W_n(z,\lambda)$ differs only
 slightly from the one corresponding to
$W_{n,\alpha}(z)$ and can be obtained from the latter SG by its small deformation.

By the above definition the Stokes lines and the corresponding Stokes graph are defined in the considered limit on the
two sheeted Riemann surface $R_2$ with the turning points of $W_{n,\alpha}(z)$ as the branch points of this surface. However
since on these two sheets the values of $\sqrt{W_{n,\alpha}(z)}$ differ by a sign only the projections on the $z$-plane
of the Stokes lines defined on each sheet coincide.

Therefore considering a pattern of SL's on the cut $z$-plane $C_{cut}$ with cuts emerging from the turning points of
$W_{n,\alpha}(z)$ we see that the SL's on $C_{cut}$ are quasi continuous on the cuts despite
the fact that they are pieces of different SL's collected from the two sheets of $R_2$.

In general, an SL emerging from a given turning point $z_i(\lambda)$ can run to infinity of $C_{cut}$
or end at another turning point $z_j(\lambda)$. An SL with the last property is called the
{\it inner} one.

An SG is called {\it critical} if at least one of its SL's is the inner one. It is called {\it
not critical} in the opposite case.

\section{Properties of Stokes graphs corresponding to the potentials $(-i\alpha z)^n$}

\hskip+2em Consider now the SG's corresponding to $W_{n,\alpha}(z)$. For this goal assume $\lambda$ to be real and positive
for a while.

For $\alpha=1$ there is a root of $W_{n,\alpha}(z)$ at $z=z_0=i$ while the remaining ones are regularly
distributed on the circle $|z|=1$ being
located at $z=z_k=ie^{i\frac{2k\pi}{n}},\;k=\pm 1,\pm 2,...,\pm[\frac{n-1}{2}]$ and at
$z=z_{\frac{n}{2}}=-i$ for even $n$, so that the
corresponding pairs of them satisfy the
relation $z_k=-{\bar z}_{-k}$ (where bar over $z$ means its complex conjugation), i.e. these pairs are located
symetrically with respect to the imaginary axis.

For $\alpha=e^{-i\frac{\pi}{n}}$ all zeros are given by pairs
$z_k=e^{i\frac{2k\pi}{n}},\;k=\pm 1,\pm 2,...,\pm\frac{n}{2}$.

It can be shown by a direct calculation (see Appendix 1 and \cite{6}) that
\be
\int_{z_{k}}^{z_{-k}} \sqrt{W_{n,\alpha}(z)}dz=\ll\{\ba{cllrc}
\frac{2i}{n}\sin\frac{2k\pi}{n}B\ll(\frac{3}{2},\frac{1}{n}\r)&for&\alpha=1,&k=1,...,&k<\frac{n}{2}\\
\frac{2i}{n}\sin\frac{(2k-1)\pi}{n}B\ll(\frac{3}{2},\frac{1}{n}\r)&
for&\alpha=e^{-\frac{i\pi}{n}},&k=1,...,&k\leq\frac{n}{2}
\ea\r.
\label{4a}
\ee
where $B(x,y)$ is the betha function.

Therefore
\be
\Re\int_{z_{k}}^{z_{-k}} \sqrt{W_{n,\alpha}(z)}dz = 0,\;\;k=1,2...,
\label{5}
\end{eqnarray}
for any such a pair, i.e. each pair $z_{k},z_{-k}$ lies on the same Stokes line.

Let us remind further that from each root of $W_{n,\alpha}(z)$ (all the roots are simple) emerge three SL's. If these
roots are members of a pair $z_{k},z_{-k},\;k=1,2...,$ then one of this lines runs from one root to another while the
remaining pairs of SL's run to infinity of the $z$-plane.

If a root of $W_{n,\alpha}(z)$ is not a member of any pair of them then all three SL's which emerge of it run to the
infinity of the $z$-plane.

Each pair of neighbour SL's emerging from the same root and running to the infinity forms a {\it sector} while the SL's alone
lie on its boundary.

However for an odd $n$ there is still a sector which boundary is formed by the neighbouring SL's runnig to infinity and
emerging from the last pair (with the highest value of $k=\frac{n-1}{2}$) of roots and by the SL linking this pair.

It is easy to note that $2n+4$ is the total number of sectors lying on $R_2$. However because of coincidence of SL's
projected on $C_{cut}$ we can consider on this cut $z$-plane quasi sectors formed by these projected SL's. There are now
$n+2$ of such quasi sectors which will be called again sectors for simplicity. We can enumerate them correspondingly to
roots attached to their boundaries.

So for each pair of turning points $z_{k},z_{-k},\;k=1,2...,$ the corresponding sectors are denoted by $S_k$ and $S_{-k}$
respectively while the single sector formed by the last pair of roots in the odd-$n$
case is denoted by $S_{\frac{n+1}{2}}$ and by $S_{\frac{n+2}{2}}$ in the even-$n$ case and in the same case the single
sector formed by the first pair of roots is denoted by $S_0$.

If there are single roots, at $z=z_0=i$ or $z=z_\frac{n}{2}=-i$, then the two sectors connected with the first root we denote by
$S_{\frac{n+3}{2}}$ (the left one) and $S_{-\frac{n+3}{2}}$ for $n$ odd and by $S_{\frac{n+2}{2}}$ (the left one) and
$S_{-\frac{n+2}{2}}$ for $n$ even. The second root at
$z_\frac{n}{2}$ can exists only for $n$ even and the two sectors connected with it are denoted by $S_{\frac{n}{2}}$
(the left) and $S_{-\frac{n}{2}}$.

Therefore typical SG's for the cases considered look as in Fig.1a-e.

\begin{tabular}{c}
\psfig{figure=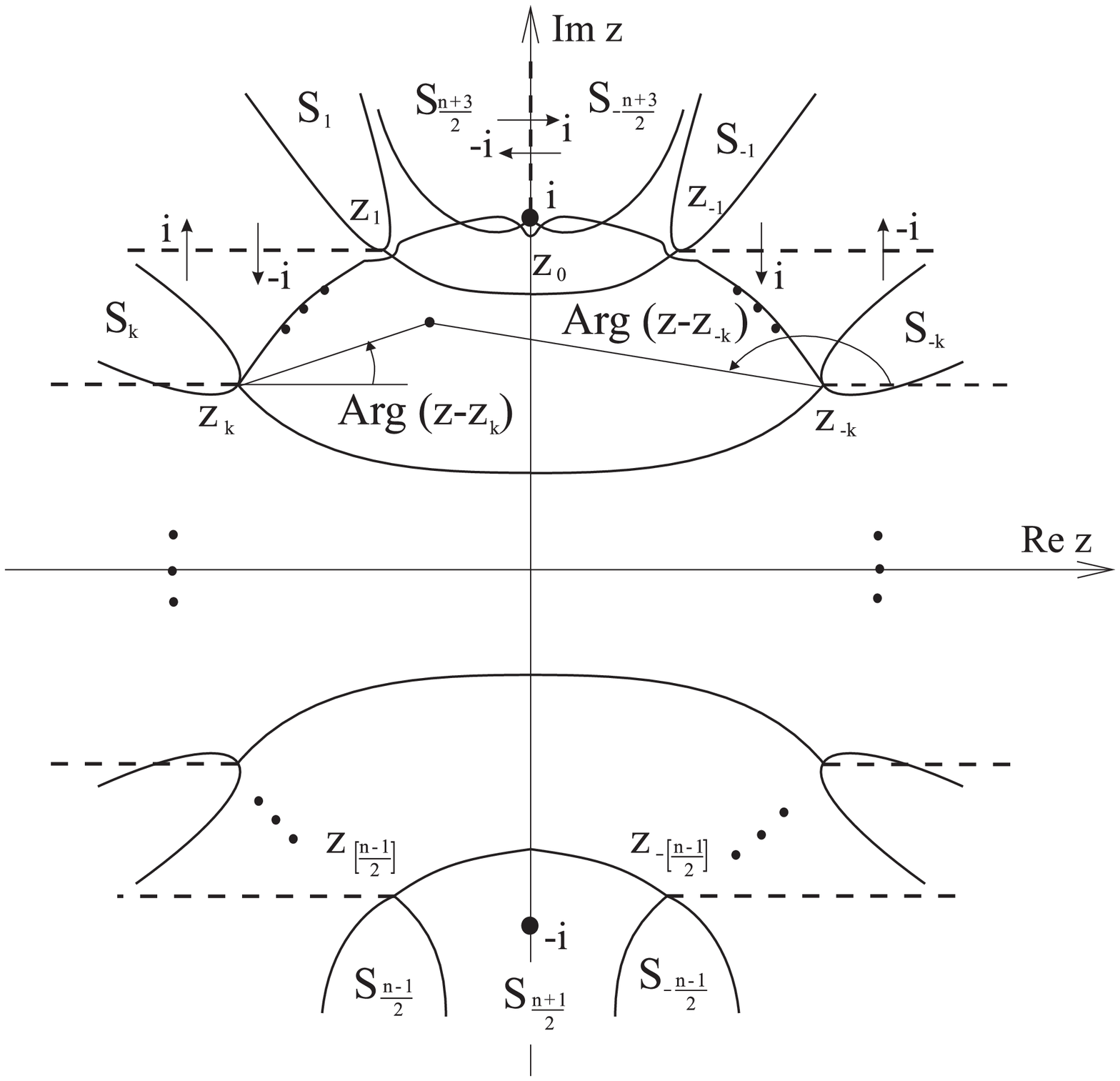,width=11cm}\\
Fig.1a SG for an odd $n$. Broken lines are cuts
\end{tabular}

\vskip 12pt

Now consider SG's corresponding to not standard cases of zeros distributions, i.e. when the standard distributions are
rotated by some angle $\phi,\;0<\phi<\frac{\pi}{n}$. Then the corresponding Stokes line linking a
pair $z_{i}',z_{j}',\;0\leq\arg z_i'<\arg z_j'\leq 2\pi,$
of not neighbouring roots of $W_{n,\alpha}(ze^{i\phi})$ is determined by the equation
\be
\Re\ll(\lambda\int_{z_{i}'}^{z_{j}'} \sqrt{W_{n,\alpha}(ze^{i\phi})}dz\r) = 0
\label{6}
\end{eqnarray}

Making a rotation $z\to ze^{i\omega}$ where $\omega=\fr(\arg z_i'+\arg z_j'\pm\pi)$ (we have to choose one
of the possibilities) we come back to the standard configuration
of roots of $W_{n,\alpha'}(z)$ (perhaps with the different $\alpha'$) and the condition \mref{6} reads then
\be
\Re\ll(\lambda e^{i\omega}\int_{z_k}^{z_{-k}} \sqrt{W_{n,\alpha'}(z)}dz\r) = 0
\label{7}
\end{eqnarray}
where $z_k,z_{-k}$ is a pair of roots of $W_{n,\alpha'}(z)$ into which the pair $z_{i}',z_{j}'$ is transformed after the
$\omega$-rotation.

Taking now $\arg\lambda=-\omega$ we can satisfy \mref{7} according to \mref{5}.

We can conclude therefore that for an arbitrary not standard distribution of roots a not neighbouring pair of them can be
connected by a Stokes line
by a proper choice of argument of $\lambda$ and if this choice is made we can cosider the arising Stokes graph in its
standard position by making the corresponding $\omega$-rotation.

\begin{tabular}{c}
\psfig{figure=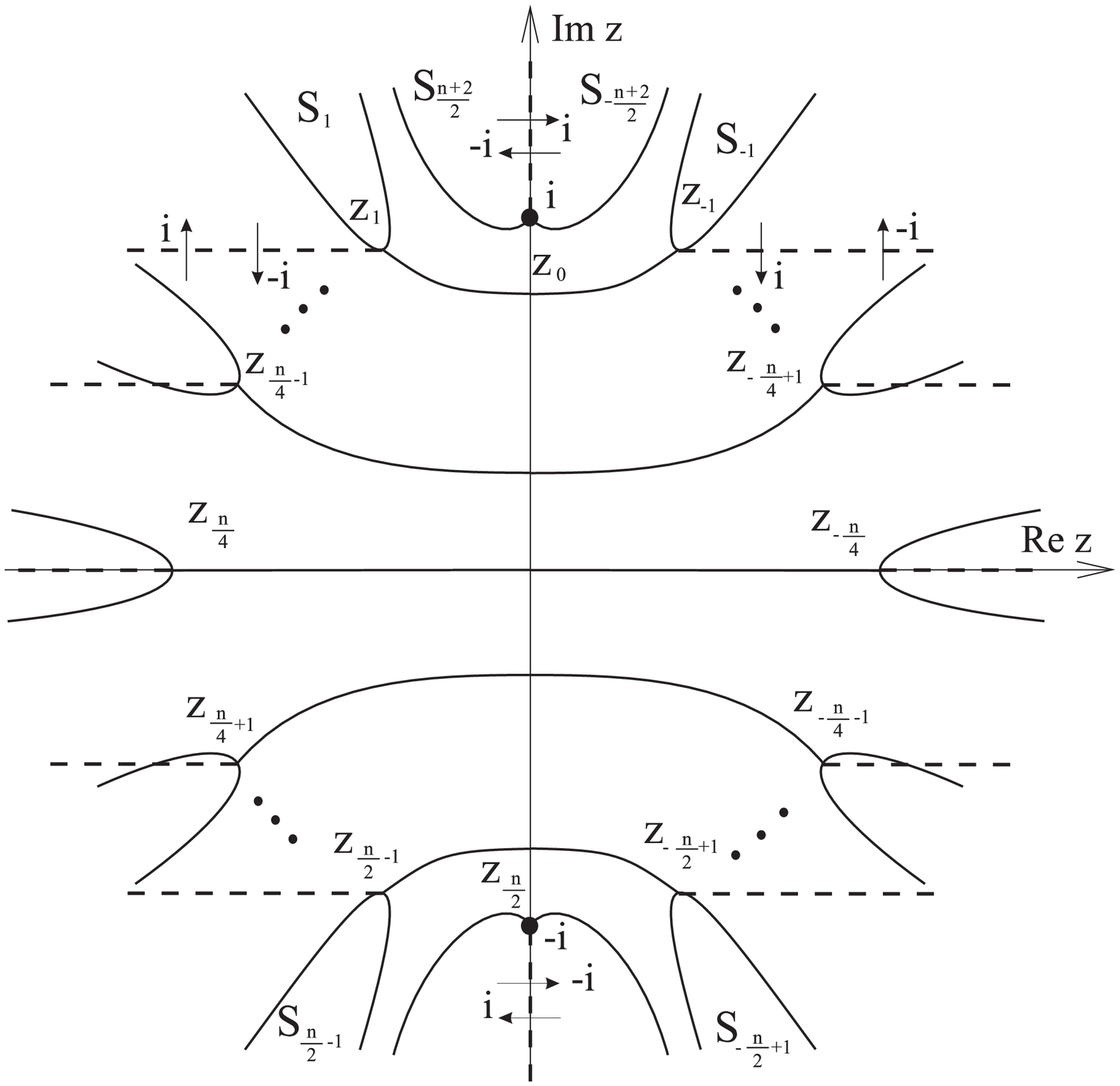,width=11cm}\\
Fig.1b SG for an even $n$ and $\alpha=1$. Broken lines are cuts
\end{tabular}

\vskip 12pt

We have got even more. Namely, if the condition \mref{6} is satisfied for a pair $z_{i}',z_{j}'$ then it is satisfied also
for all pairs of turning points which are symmetric
with respect to the line $\arg z=\fr(\arg z_{i}'+\arg z_{j}')$ since the corresponding $\omega$-rotations are the same
for such pairs and coincide with the one for the pair $z_{i}',z_{j}'$.

Consider now again the SG's corresponding to the standard configurations of roots and for real $\lambda$. They are shown on
Fig.1a-e. What happens when $\lambda$ acquire a nonvanishing phase $\beta$? It is well known that in such a case each three
SL's emerging from the same root rotate by an angle $-\frac{2\beta}{3}$ and of course for $\beta=\pm \pi$ the whole SG
comes back to its form for real $\lambda$.

However patterns of SG's arising by this $\beta$-rotations are not all topologically different because of symmetric
distributions of turning points. Namely, if we consider the SL linking a pair $z_{k},z_{-k}$ (call it $L_k$) and start to rotate the SG
by sufficiently small $\beta >0$ then, since all SL's of the graph will rotate anticlockwise, $L_k$ will split into two
SL's now runnig to the infinity each. Still enlarging $\beta$ we can cause one of these splitting SL's emerging from $z_k$
to meet the root $z_{-k-1}$, i.e. the two last roots find themselves on the same SL. But this is just the situation we
have discussed previously so that by the corresponding $\omega$-rotation in the $z$-plane we can return to the standard configuration of
SG. It is easy to see that for this case the $\omega$-rotation has to be done by one of the angles
$\pm\frac{\pi}{n},\pi\pm\frac{\pi}{n}$. A choice of the two latter angles corresponds to an odd $n$ only and has to be done
only when really the SL arises between the roots $z_k$ and $z_{-k-1}$.

\begin{tabular}{c}
\psfig{figure=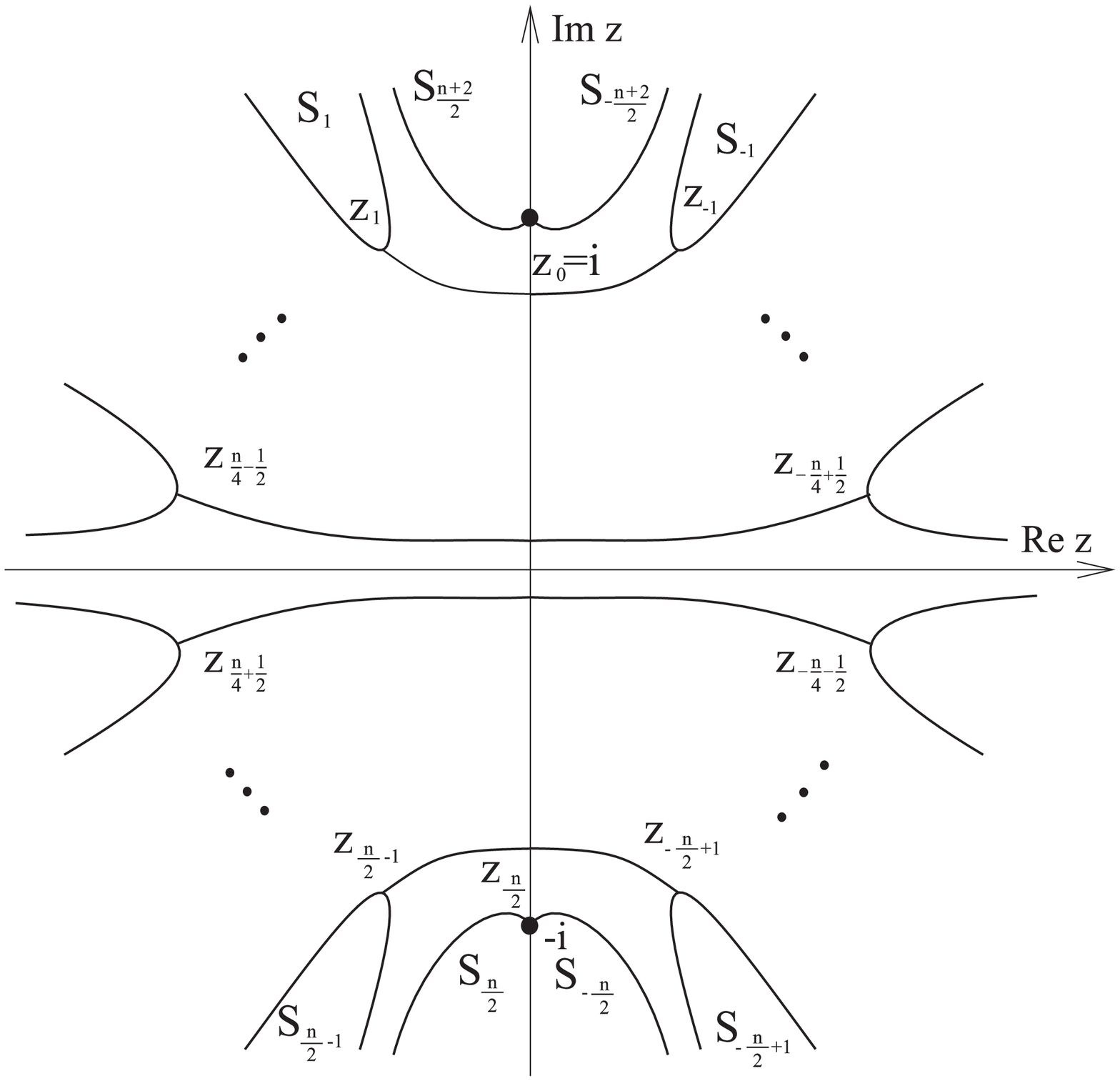,width=11cm}\\
Fig.1c SG for an even $n$ and $\alpha=1$
\end{tabular}

\vskip 12pt

Therefore as it follows from the above discussion it is enough to rotate the standard SG by angles $\beta$ chosen from the
interwals $(-\frac{\pi}{n},\frac{\pi}{n})$ to handle all topologically nonequivalent configurations of SG for the
standard distributions of roots.

Let us put therefore $\lambda=|\lambda|e^{i\beta},\;0<|\beta|<\frac{\pi}{n}$, in the condition
\be
\Re\ll(\lambda \int_{z_k}^z \sqrt{W_{n,\alpha}(\xi)}d\xi\r) = 0
\label{8}
\end{eqnarray}
where $k$ takes all possible values following from the values of $n,\;n=2,3,4,...$ .

It then follows from the above discussion that all SG's corresponding to the $\lambda$'s chosen are non-critical, i.e.
each of their SL's defined by \mref{8} runs to the infinity of the $z$-plane, see Fig.2.

We shall show in the next sections that for the potentials  $(-i\alpha z)^n$ and $W_n(z)+1$
in the limit $|\lambda|\to\infty$ all zeros of each of their fundamental solutions are distributed along
a boundary of its canonical domain, while this boundary is a collection of Stokes lines.

\begin{tabular}{c}
\psfig{figure=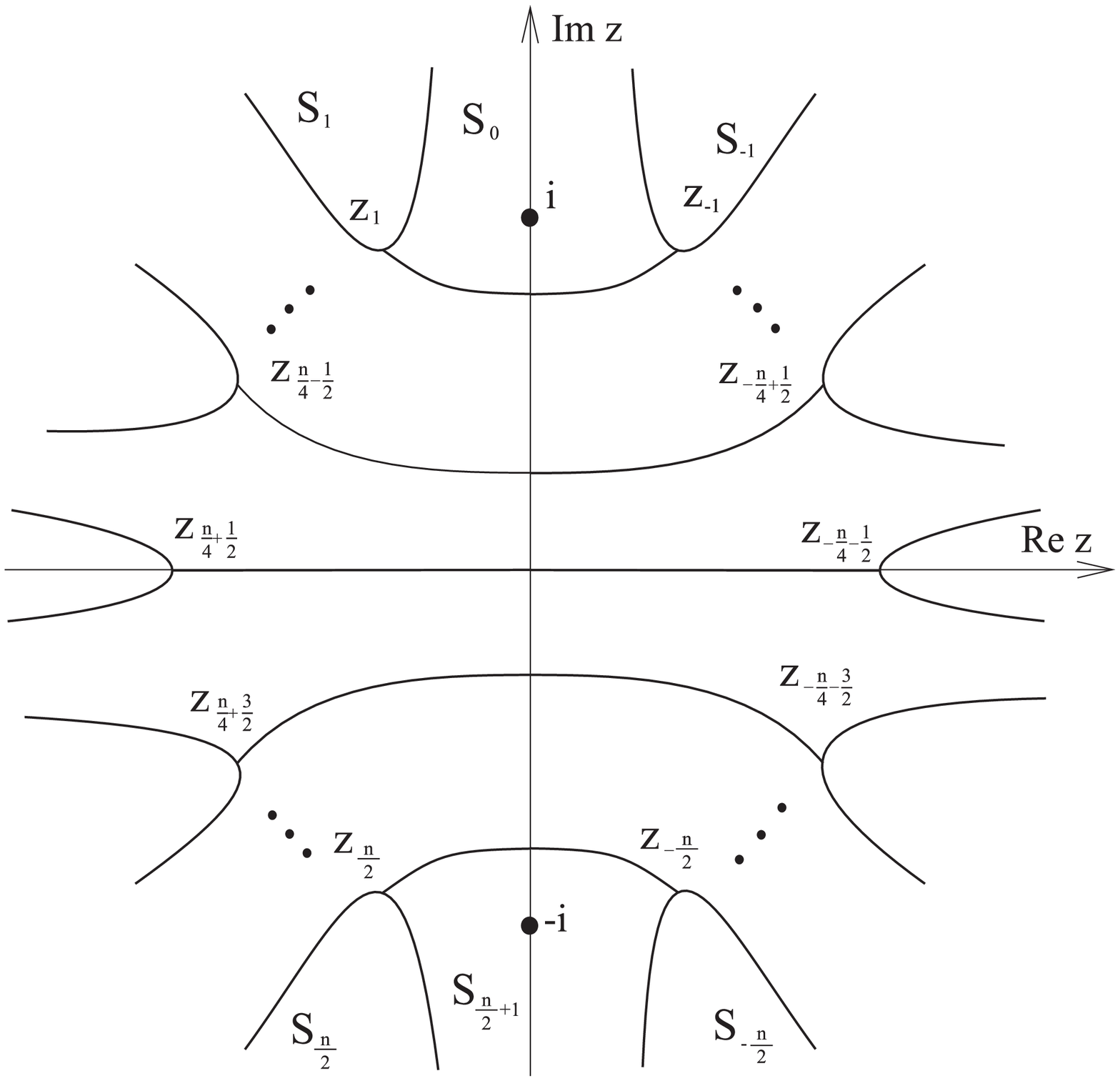,width=11cm}\\
Fig.1d SG for an even $n$ and $\alpha=e^{-\frac{i\pi}{n}}$
\end{tabular}

\vskip 12pt

\section{Fundamental solutions for the potential $(-i\alpha z)^n$}

\hskip+2em The fundamental solutions (FS) of the equation
\be
\psi''(z)-\lambda^2\ll((-i\alpha z)^n-1\r)\psi(z)=0
\label{9}
\ee
are solutions defined on $R_2$ separately in each of $2n+4$ sectors. They are
subdominant in the sectors which they are defined in, i.e.
they vanish for $|z|\to\infty$ inside the sectors. They can be given explicite forms of the following functional series
\cite{3}:
\be
\psi_{\alpha,k}(z,\lambda) = W_{n,\alpha}^{-\frac{1}{4}}(z)e^{\sigma_k\lambda {\tilde W}_{n,\alpha}(z,z_k)}
\chi_{\alpha,k}(z,\lambda)
\label{10}
\end{eqnarray}
where $z\in S_k$ and $z_k$ is a turning point lying on the boundary of $S_k$ while
\begin{eqnarray}
{\tilde W}_{n,\alpha}(z,z_k)=\int_{z_k}^z \sqrt{W_{n,\alpha}(y)}dy\nn\\
\chi_{\alpha,k}(z,\lambda) = 1 + \sum_{n{\geq}1}
\left(-\frac{\sigma_k}{2\lambda} \right)^{n}Y_{\alpha,k;n}(z,\lambda)
\label{11}
\ee

\begin{tabular}{c}
\psfig{figure=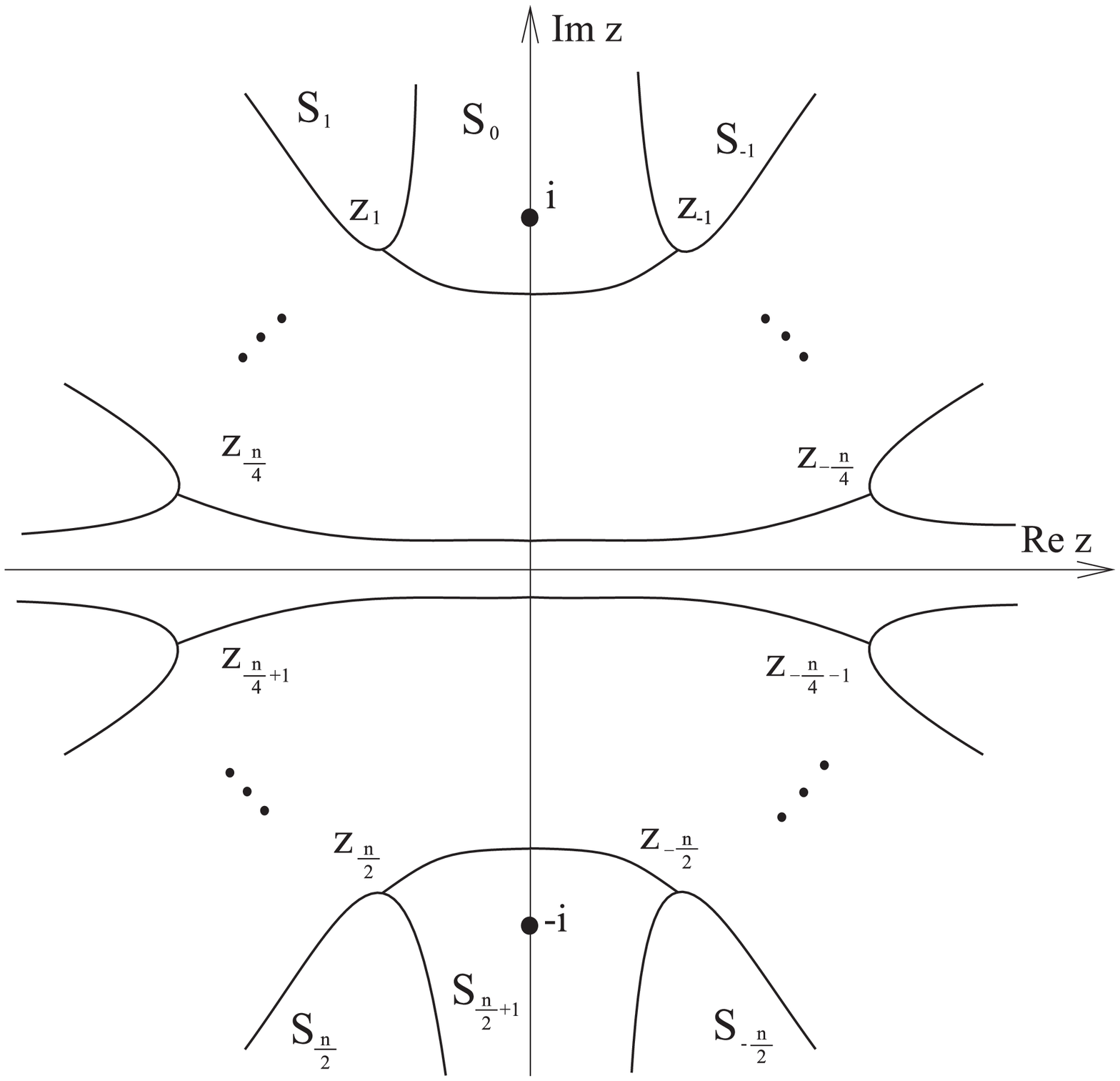,width=11cm}\\
Fig.1e SG for an even $n$ and $\alpha=e^{-\frac{i\pi}{n}}$
\end{tabular}

\vskip 12pt

and
\be
 Y_{\alpha,k;n}(z,\lambda) =\int_{\gamma_k(z)}d{y_{1}}
\int_{\gamma_k(y_1)}d{y_{2}} \ldots
\int_{\gamma_k(y_{n-1})}d{y_{n}}
\omega_\alpha(y_{1})\omega_\alpha(y_{2}) \ldots \omega_\alpha(y_{n}){\times}\nn\\
\ll( 1 - e^{-2\sigma_k\lambda {\tilde W}_{n,\alpha}(z,y_1)} \right)
\left(1 - e^{-2\sigma_k\lambda {\tilde W}_{n,\alpha}(y_1,y_2)}\right)\cdots
\ll(1 - e^{-2\sigma_k\lambda {\tilde W}_{n,\alpha}(y_{n-1},y_n)}\right)\nn\\n\geq 1
\label{12}
\ee
with
\be
\omega_{\alpha}(z) ={\frac{5}{16}}{\frac{W_{n,\alpha}^{\prime 2}(z)}{W_{n,\alpha}^{\frac{5}{2}}(z)}}  -
{\frac{1}{4}}{\frac{W_{n,\alpha}^{\prime\prime}(x)}{W_{n,\alpha}^{\frac{3}{2}}(x)}}=
{W_{n,\alpha}^{- \frac{1}{4}}(z)} \left( {W_{n,\alpha}^{- \frac{1}{4}}(z)} \right)^{\prime{\prime}}
\label{13}
\ee

The signatures $\sigma_k=\pm 1$ present in the formulae \mref{10}-\mref{12} are defined in each particular sector $S_k$
in such a way to ensure the inequality $\Re(\sigma_k\lambda{\tilde W}_{n,\alpha}(z,z_k))<0$ to be
satisfied in this sector.

The integration paths $\gamma_k(z)$ in \mref {12} which start from the infinities of the corresponding sectors are chosen
in such a way to satisfy the inequality
$\Re(\sigma_k\lambda{\tilde W}_{n,\alpha}(y_k,y_{k+1}))\geq 0,\;y_k,y_{k+1}\in\gamma,$ for each factor of
the integrations in \mref{12}. Such paths are called canonical and a point $z$ for which $\gamma_k(z)$ can be chosen to be
canonical is also called canonical with respect to the sector $S_k$. The latter property is completely of a
topological nature.

According to our earlier discussion we can assume the argument of $\lambda$ to vary in the interval ($-\frac{\pi}{n},+
\frac{\pi}{n}$).

Let us note however that there is no necessity to consider the fundamental solutions on the whole $R_2$. It is enough to
define them and consider on $C_{cut}$ only because of the following reasons.

As we have mentioned earlier there is
one to one correspondence between the sectors lying on the $C_{cut}$ which can be considered as a one sheet of $R_2$ and
the sectors lying on the second sheet $C_{cut}'$ of $R_2$ the latter sheet being then a complement of $C_{cut}$ to $R_2$
connected with $C_{cut}$ by cuts. The correspondence between two such sectors is built by their coincidence
when $C_{cut}'$ is projected on $C_{cut}$.

Let $S\subset  C_{cut}$ and $S'\subset  C_{cut}'$ be a pair of such sectors. Then the two fundamental solutions defined
in each of these two sectors coincide up to a constant. This coincidence is visible in the form given by \mref{10}-
\mref{13} by the invariance of $\sigma_k{\tilde W}_{n,\alpha}(z,z_k)$ and
$\sigma_k\omega_\alpha(z)$ when passing from the sector $S$ to $S'$ since
then $\sigma_k$, ${\tilde W}_{n,\alpha}(z,z_k)$ and $\omega_\alpha(z)$ change their signs simultanuously. Only the common
factor $W_{n,\alpha}^{-\frac{1}{4}}(z)$ of the solutions changes slightly under this
operation acquiring one of the phase factors $\pm i$.

Therefore we can consider the fundamental solutions defined only on $C_{cut}$ and
the solution $\psi_{\alpha,k}(z,\lambda)$ given by \mref{10}-\mref{13} is defined in the sector $S_k$ the latter being
enumerated in the way described earlier. However if $\psi_{\alpha,k}(z,\lambda)$ is defined in the sector crossed by a cut of
$C_{cut}$ we have to remember about necessary changes in the forms \mref{10}-\mref{13} of the solution described above
to keep its identity when the cut is crossed. This latter note is valid also when such a cut is crossed by any of the
fundamental solutions if the latter is continued analytically along a canonical path.

A collection $D_k\subset C_{cut}$ of all points canonical with respect to $S_k$ is called a canonical domain corresponding to $S_k$. Of
course $S_k\subset D_k$ for each $k$ \cite{3}.

A general rule for a given SL to belong to $\partial D_k$ is that a canonical path $\gamma_k(z)$ when $z$ approaches the
line has to cross another SL emerging from the same turning point.

For non-critical SG's (i.e. all SL's of which run to the infinities) $\partial D_k$ is composed of all turning points
and of single SL's emerging from these turning points, see Fig.2.

In other cases (i.e. for critical positions of SG) all three SL's emerging from a turning point
can belong to $\partial D_k$. The latter point is then just the root $z_{-k}$ joined by one of its Stokes lines with
the turning point $z_k$, i.e. by the line which
start from the former point and end at the latter. Such a line is then called an {\it inner} one. All points of the sector
$S_{-k}$ cannot be then joined with the infinity of $S_k$ by cannonical paths and therefore do not belong to $D_k$.

Each $\partial D_k$ is specific for its canonical domain $D_k$.

In what follows we will consider also a $C_{cut}(\epsilon)$-plane arising from $C_{cut}$ by depriving the latter
$\epsilon$-vicinities of the turning points $z_k,\;k=1,..,n$. Namely, for any $\epsilon>0$ denote by $\Delta_k(\epsilon)$
the following circle vicinities of turning points $z_k$: $\Delta_k(\epsilon)=\{z:|z-z_k|<\epsilon\},\;k=1,...,n$. Then
$C_{cut}(\epsilon)=C_{cut}\setminus\bigcup_{k=1}^n{\bar \Delta}_k(\epsilon)$.

Let $L_k^r$ denotes a connected set of SL's contained in $\partial D_k$ and emerging from the turning points $z_r$. Therefore
for the non-critical SG's all $L_k^r$ contain only single SL's and are disjoint pairways while for the critical SG's
a unique difference with the previous case is connected with $L_k^k$ and $L_k^{-k}$ which are not disjoint, the first one containing
the inner SL between $z_k$ and $z_{-k}$ while the second one containing all the three SL's emerging from $z_{-k}$,
therefore also the inner SL between $z_k$ and $z_{-k}$. The lines $L_k^r,\;r=1,...,n$, will be called (after Eremenko
{\it et al} \cite{8}) {\it exceptional} with respect to the solution
$\psi_{\alpha,k}(z)$. Of course they are exceptional in that none of their points can be reached by the solution
$\psi_{\alpha,k}(z)$ if the latter is to be continued to them along canonical paths.

For a given $L_k^r$ let us denote by $V_k^r(\epsilon)$ an $\epsilon$-vicinity of this set of SL's defined by the following
conditions:
\begin{enumerate}
\item $L_k^r\subset V_k^r(\epsilon)$
\item a boundary of $V_k^r(\epsilon)$ consists of (at most two) continues lines an Euclidean
distance of which to $L_k^r$ is smaller than $\epsilon$ (see Fig.2 and 3.)
\end{enumerate}

Let us further denote by $D_{k,\epsilon}$ a subset of $D_k$ given by
$D_{k,\epsilon}=D_k\setminus{\bar V}_k(\epsilon)$ where $ V_k(\epsilon)=\bigcup_{r=1}^n V_k^r(\epsilon)$.

It is clear that $V_k(\epsilon)$ is an $\epsilon$-vicinity of $\partial D_k$ (see Fig.2 and 3.)

\vskip 18pt

\begin{tabular}{c}
\psfig{figure=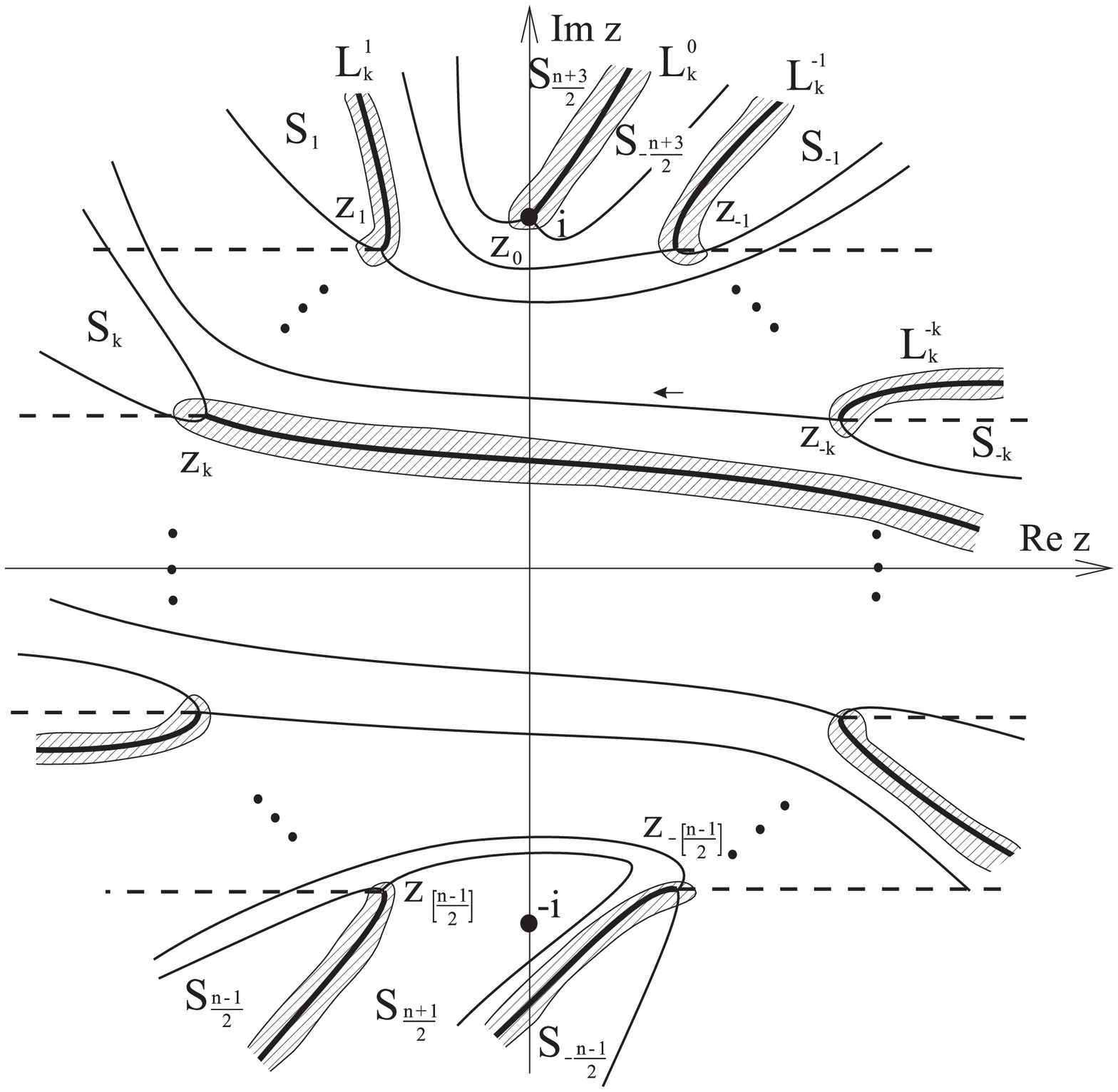,width=11cm}\\
Fig.2 $V_{k,\epsilon}$-vicinity of a domain where $\psi_k(z)$ cannot be continued canonically.\\ The non-critical case
\end{tabular}

\vskip 12pt

Needless to say for the non-critical SG's every turning point $z_l$ belongs to $V_k(\epsilon)$ together with
exactly one of its three SL's which emerge from it. In the case of the critical SG's the same is true for all
$z_l$ excluding $z_{-k}$ which belong to $V_k(\epsilon)$ together with its three SL's if $z_k$ is connected with $z_{-k}$
by the inner SL.

The following property of FS's is the key one for our further considerations.

\vskip 12pt

{\bf Lemma}

{\it In the domain $D_{k,\epsilon}$ the factor $\chi_{\alpha,k}(z,\lambda)$ of the solution} \mref{10} {\it satisfies the following bound
\be
|\chi_{\alpha,k}(z,\lambda)-1|\leq e^\frac{C_{\alpha;\epsilon}}{\lambda_0}-1,\;\;\;\;\;|\lambda|>\lambda_0\nn\\
C_{\alpha;\epsilon}=\liminf_{\gamma_k(z),\;z\in D_{k,\epsilon},\;k=1,...,n}\int_{\gamma_k(z)}|\omega_\alpha(\xi)d\xi|<\infty
\label{14}
\ee
where $\gamma_k(z)$ are canonical}.

We have left the proof of {\bf Lemma} to Appendix 2.

\begin{tabular}{c}
\psfig{figure=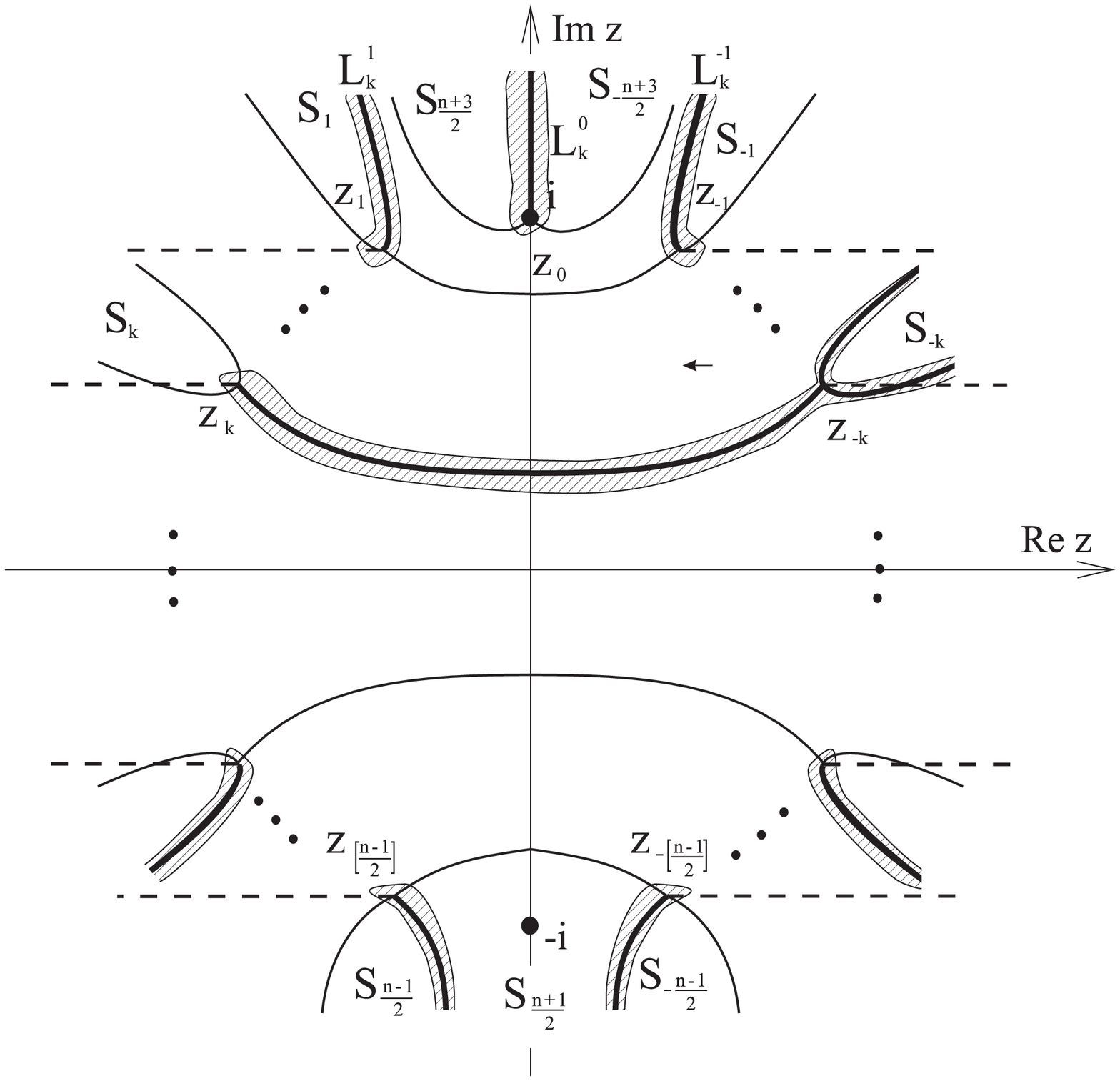,width=11cm}\\
Fig.3 $V_{k,\epsilon}$-vicinity of a domain where $\psi_k(z)$ cannot be continued canonically.\\ The critical case
\end{tabular}

\vskip 12pt

There is a direct relation between above {\bf Lemma} and so called semiclassical expansions of $\chi$-factors of FS's
\mref{10} for $\lambda\to\infty$ but fixed $z$ which will be needed in our further considerations. Such
expansions have been considered in our earlier papers (see \cite{4}, ref.1 and \cite{5}, ref.5) and have been given the
following exponential forms:
\be
\chi_{\alpha,k}(z,\lambda)\sim\chi_{\alpha,k}^{as}(z,\lambda)=1 + \sum_{p{\geq}1}
\left(-\frac{\sigma_k}{2\lambda} \right)^{p}{\tilde Y}_{\alpha,k;p}(z)=
\sum_{p{\geq}0}\left(-\frac{\sigma_k}{2\lambda} \right)^{p}{\tilde Y}_{\alpha,k;p}(z)=\nn\\
\exp\ll(\sum_{p{\geq}1}\left(-\frac{\sigma_k}{2\lambda}\right)^p\int_{\infty_k}^zX_{\alpha,p}(y)dy\r)
\label{551}
\ee
where
\be
{\tilde Y}_{\alpha,k;0}(z)\equiv 1\nn\\
{\tilde Y}_{\alpha,k;p}(z)=\int_{\infty_k}^zdy_nW_{n,\alpha}^{-\frac{1}{4}}(y_p)\ll(W_{n,\alpha}^{-\frac{1}{4}}(y_n)
\int_{\infty_k}^{y_p}dy_{p-1}W_{n,\alpha}^{-\frac{1}{4}}(y_{p-1})\times\r.\nn\\
\ll(...\ll.W_{n,\alpha}^{-\frac{1}{4}}(y_2)\int_{\infty_k}^2dy_1W_{n,\alpha}^{-\frac{1}{4}}(y_1)
\ll(W_{n,\alpha}^{-\frac{1}{4}}(y_1)\r)''...\r)''\r)'',\;\;\;p\geq 1
\label{552}
\ee
and
\be
\sum_{p{\geq}1}\left(-\frac{\sigma_k}{2\lambda}\right)^pX_{\alpha,p}(z)\equiv Z_{\alpha,k}(z,\lambda)=
\frac{1}{\chi_{\alpha,k}^{as}(z,\lambda)}\frac{d\chi_{\alpha,k}^{as}(z,\lambda)}{dz}
\label{553}
\ee

Note that $X_{\alpha,p}(y),\;p\geq 1$, are sector independent being point functions on the
$C_{cut}$-plane which are given by the following recurrent formula (see \cite{4}, ref.2):
\be
X_{\alpha,1}(z)=\omega(z)=W_{n,\alpha}^{-\frac{1}{4}}(z)\ll(W_{n,\alpha}^{-\frac{1}{4}}(z)\r)''=
W_{n,\alpha}^{-\fr}(z)\frac{U_{2n-2}(z)}{W_{n,\alpha}^2(z)}\nn\\
X_{\alpha,p}(z)=-\fr W_{n,\alpha}^{-\frac{3}{2}}(z)W_{n,\alpha}'(z)X_{\alpha,p-1}(z)+\nn\\
W_{n,\alpha}^{-\frac{1}{2}}(z)\ll(\sum_{k=1}^{m-2}X_{\alpha,k}X_{\alpha,m-k-1}+X_{\alpha,p-1}'(z)\r),\;\;\;\;\;
p=2,3,...
\label{554}
\ee
where $U_{2n-2}(z)$ is a polynomial of the $2n-2$-degree.

It follows from \mref{554} that $X_{\alpha,2m},\;m\geq 1,$ have only poles at the turning points while
$X_{\alpha,2m+1},\;m\geq 0,$
have there the square root branch points. Therefore the same are the properties of $Z_\alpha^+(z,\lambda)$ and
$Z_\alpha^-(z,\lambda)$
at these points respectively where $Z_\alpha^+(z,\lambda)+\sigma_kZ_\alpha^-(z,\lambda)=Z_{\alpha,k}(z,\lambda)$ and
\be
Z_\alpha^+(z,\lambda)=\sum_{m{\geq}1}\left(\frac{1}{2\lambda}\right)^{2m}X_{\alpha,2m}(z)\nn\\
Z_\alpha^-(z,\lambda)=\sum_{m{\geq}0}\left(\frac{1}{2\lambda}\right)^{2m+1}X_{\alpha,2m+1}(z)
\label{555}
\ee

If we now take into account that $\chi_{\alpha,i\to j}(\lambda)\equiv\lim_{z\to\infty_j}\chi_{\alpha,i}(z,\lambda)=
\chi_{\alpha,j\to i}(\lambda)$ where $\infty_j$ is the infinite point of the sector $S_j$ communicated canonically with the sector $S_i$
(see  ref.1 of \cite{4} and ref.5 of \cite{5}) then we get
\be
e^{\int_{\infty_i}^{\infty_j}(Z_\alpha^++\sigma_iZ_\alpha^-)dz}=e^{\int_{\infty_j}^{\infty_i}(Z_\alpha^++\sigma_jZ_\alpha^-)dz}
\label{556}
\ee

Since however $\sigma_i=-\sigma_j$ then we get from \mref{556}
\be
\int_{\infty_i}^{\infty_j}Z_\alpha^+(z,\lambda)dz=0
\label{557}
\ee
for any pair of canonically communicated sectors.

However the integration in \mref{557} is now not limited by canonical paths since under the integral there are now no
exponentials limiting this integration to canonical paths, i.e. these paths can be freely deformed with the integral
beeing still
convergent. It is easy to see that because of that a given integral \mref{557} can be deformed to any other integral
between any pair of
sectors (i.e. not necessarily communicated canonically) as well as to any integral along an arbitrary loop. It means that
residua of $Z_\alpha^+(z,\lambda)$ at the poles which it has at the turning points vanish.

Therefore we conclude that the Riemann surface of $Z_\alpha^+(z,\lambda)$ is just the $C$-plane on which it is meromorphic with
vanishing residua at its poles. It means of course that this is the property of each $X_{\alpha,2m},\;m\geq 1$ as well. Thus
when integrating $Z_{\alpha,k}(z,\lambda)$ along contours starting and ending at the same points we get only contribution from
the odd part of $Z_{\alpha,k}(z,\lambda)$, i.e. from $\sigma_kZ_\alpha^-(z,\lambda)$.

\section{High energy limit of loci of zeros of fundamental solutions for the potential $(-i\alpha z)^n$}

\hskip+2em It follows from the {\bf Lemma} that making $\lambda_0$ sufficiently large, i.e. $\lambda_0\gg C_{\epsilon}$, we
can
make $\chi_{\alpha,k}(z,\lambda)$
arbitrarily close to unity for $z\in D_{k,\epsilon}$ and $|\lambda|>\lambda_0$. But this means that for such conditions
$\psi_{\alpha,k}(z,\lambda)$ vanishes nowhere in $D_{k,\epsilon}$. Therefore, we have
the following theorem about loci of zeros of $\psi_{\alpha,k}(z,\lambda)$.

{\bf Theorem 1}

{\it For sufficiently large $\lambda$ all zeros of $\psi_{\alpha,k}(z,\lambda)$ can lie exclusively in the completion $C_{cut}\setminus D_{k,\epsilon}$
of the domain $D_{k,\epsilon}$.}

In the case of
a non-critical SG this completion coincides of course with $V_{k,\epsilon}$ while in the critical case it can contain also
a whole
sector. This is because in the case of the potential considered a
connected set of SL's which contains more than three lines with three of them emerging from the same turning point $z_k$
has to contain also one more
turning point, i.e. $z_{-k}$. If we consider a solution $\psi_{\alpha,k}(z,\lambda)$ defined in the sector $S_k$ then the sector
$S_{-k}$ and its boundary $\partial S_{-k}$ cannot be then connected with $S_k$ by canonical paths, i.e. they lie in the
completion $C_{cut}\setminus D_k$. We see that $S_{-k}$ is just the sector which has been mentioned earlier as
contained in $C_{cut}\setminus D_{k,\epsilon}$ (see Fig.3).

To formulate a theorem giving the precise positions of the roots of $\psi_{\alpha,k}(z,\lambda)$ mentioned in {\bf Theorem 1} for
the cases of non-critical SG's let us make the following several notes.

The first one is that in choosing $C_{cut}$ we can always choose
the corresponding cuts in
such a way to ensure all the SL's lying in $C_{cut}$ to be strictly (i.e. not quasi) continuous in this cut plane, i.e. none of
these lines can cross any cut. With such a choice for every root $z_k$ of $W_{n,\alpha}(z)$ the values of
$\Im\ll(\lambda\int_{z_{k}}^{z} \sqrt{W_{n,\alpha}(y)}dy\r)$
have definite sign on every SL emerging from $z_k$ depending on the line.

The second one is, as we have already mentioned, that in the case of non-critical SG's every
turning point $z_l$ belongs to $\p D_k$
together with exactly one of its three SL's which emerge from it and $\p D_k$ is
collected
exactly of only such SL's.

The third one is the most important since it establishes the way of taking the limit $\lambda\to\infty$. Namely, as it
follows from App.2 zeros of $\psi_{\alpha,k}(z,\lambda)$ scattered around such an exeptional line sliding down in the limit
considered in directions to the respective turning point from which the line emerges. If these zeros are not sufficiently
far from the turning point they unavoidably approach their limit just in this point. However in such a case the conditions
of the validity of the representations \mref{10}-\mref{13} and their semiclassical expansions \mref{551}-\mref{554} are
broken, i.e. the limit positions of zeros are outside of the domains $D_{k,\epsilon}$ engaged in taking the limit. A
necessary condition to get the limit loci of zeros different from the turning point is to take their
initial loci to be $\lambda$-dependent.
This dependence in the formulae given below is not only necessary but also sufficient
giving finite loci of zeros different from turning points.

This dependence however defines also the way of getting the limit
$\lambda\to\infty$ inforcing to put $|\lambda|=[|\lambda|]+\Lambda,\;0\leq\Lambda<1$, where $[|\lambda|]$ is a step function
of $|\lambda|$ (i.e. an integer not greater than $|\lambda|$ itself) and to consider the limit $\lambda\to\infty$ conditioned by the fixed value of $\Lambda$. This kind of limit
will be called {\it regular} to distinguish it from the {\it free} limit, i.e. with no conditions. Obviously, each
$\Lambda$ defines a different way of getting the limit $\lambda\to\infty$ by $\psi_{\alpha,k}(z,\lambda)$ and its zeros
as well.

Nevertheless we can also consider zeros the limit loci of which are just turning points when $\lambda\to\infty$. However
we have to remeber in such cases that the asymptotic formulae used to get these loci have to be bounded to the domains
$D_{k,\epsilon}$.

The following theorems give the precise limit positions of roots of
$\psi_{\alpha,k}(z,|\lambda|e^{i\beta})$ for $|\lambda|\to\infty$ up to {\it all} terms of $\lambda^{-1}$.

{\bf Theorem 2a}

{\it Zeros $\zeta_{l,qr}^{(k)}(\lambda),\;|\lambda|=[|\lambda|]+\Lambda,\;l=0,\pm 1,\pm 2,...,\;q=0,1,2,...,\;
r=0,\pm 1,\pm 2,...,$ of
$\psi_{\alpha,k}(z,|\lambda|e^{i\beta}),\;0<|\beta|<\frac{\pi}{n}$, i.e. in the
non-critical cases, in the regular limit $[|\lambda|]\to\infty$, are
distributed on $C_{cut}$ uniquely along the corresponding exceptional SL's according to the formulae}:
\be
\int_{K_l(\zeta_{l,qr}^{(k)}(\lambda))}\ll(\fr\sqrt{W_{n,\alpha}(y)}-
\frac{1}{2\lambda}Z_{\alpha,k}(y,\lambda)\r)dy=\pm\ll(q[|\lambda|]+r-\frac{1}{4}\r)\frac{i\pi}{\lambda}
\label{14a}
\ee
{\it where $K_l(\zeta_{l,qr}^{(k)}(\lambda))$ is a contour which starts and ends at
$\zeta_{l,qr}^{(k)}(\lambda)$ rounding the turning point $z_l$ anticlockwise.

Zeros $\zeta_{l,qr}^{(k)}(\lambda)$ for $q>0$ have the following semiclassical expansion:}
\be
\zeta_{l,qr}^{(k)}(\lambda)=\sum_{p\geq 0}\frac{1}{\lambda^p}\zeta_{l,qrp}^{(k)}(\Lambda)
\label{141}
\ee
{\it with two first terms given by}:
\be
\int_{K_l(\zeta_{l,qr0}^{(k)})}\fr\sqrt{W_{n,\alpha}(y)}dy=
\int_{z_l}^{\zeta_{l,qr0}^{(k)}}\sqrt{W_{n,\alpha}(y)}dy=\pm qi\pi e^{-i\beta}\nn\\
\zeta_{l,qr1}^{(k)}(\Lambda)=\pm(r-q\Lambda-\frac{1}{4})\frac{i\pi e^{-i\beta}}{\sqrt{W_{n,\alpha}(\zeta_{l,qr0}^{(k)})}}
\label{142}
\ee

{\it For $q=0$ we have instead $\zeta_{l,0r0}^{(k)}(\Lambda)\equiv z_l$ and}
\be
\int_{z_l}^{z_l+\zeta_{l,0r1}^{(k)}(\Lambda)/\lambda}\sqrt{W_{n,\alpha}(y)}dy=
(r-\frac{1}{4})\frac{i\pi}{\lambda}\nn\\
r>\frac{|\lambda|}{\pi}\limsup_{|\phi|\leq\pi}\ll|\int_{z_l}^{z_l+\epsilon e^{i\phi}}\sqrt{W_{n,\alpha}(y)}dy\r|
\label{142b}
\ee
{\it as well as}
\be
\zeta_{l,0r2}^{(k)}(\Lambda)=\frac{1}{8}\frac{\int_{K_l(z_l+\zeta_{l,0r1}^{(k)}(\Lambda)/\lambda)}X_{\alpha,1}(y)dy}
{\sqrt{W_{n,\alpha}(z_l+\zeta_{l,0r1}^{(k)}(\Lambda)/\lambda)}}
\label{142c}
\ee

{\it The signs $\pm$ above are to be chosen the same as of
$\Im\ll(e^{i\beta}\int_{z_{l}}^{z} \sqrt{W_{n,\alpha}(y)}dy\r)$ on the corresponding exceptional lines}.

First of all what follows from the formulae \mref{141} and \mref{142} is that, according to \mref{4},
$\zeta_{l,qr0}^{(k)},\;q=1,2,...,$
lie on SL emerging from $z_l$ while the same can not be said about the first term of the expansion
\mref{141}, i.e. the integral $\int_{z_{l}}^{\zeta_{l,qr}^{(k)}(\lambda)} \sqrt{W_{n,\alpha}(y)}dy$ is not
purely imaginary if the first and next orders of $\lambda^{-1}$ are included in the expansion \mref{141}.
It means that zeros $\zeta_{l,qr}^{(k)}(\lambda)$ deviate from SL considered when higher order terms
in \mref{141} are included beginning from the first one.

We get therefore the following picture of loci of zeros $\zeta_{l,qr}^{(k)}(\Lambda)$ in the considered
regular limit. For each $\Lambda,0\leq\Lambda<1$, these zeros are distributed along the SL emerging from $z_l$. For
a given $q,\;q=1,2,...,$ every two neighbouring zeros are separated from each other by
$i\pi e^{-i\beta}/\ll(\lambda\sqrt{W_{n,\alpha}(\zeta_{k,m0}^{(0)})}\r)$. For a given $q,\;q=1,2,...,$ and
$r,\;r=0,\pm 1,\pm 2,...,$ when $\Lambda$ changes from zero to unity these zeros ocupy a line:
\be
z=\zeta_{l,qr0}^{(k)}+
(r-q\Lambda-\frac{1}{4})\frac{i\pi e^{-i\beta}}{\sqrt{W_{n,\alpha}(\zeta_{l,qr0}^{(k)})}}\frac{1}{[|\lambda|]+\Lambda}
\label{142a}
\ee
which crosses the exceptional SL considered at $z=\zeta_{l,qr0}^{(k)}$ for $\Lambda=(r-\frac{1}{4})/q$.

On the other hand for higher $q=1,2,...,$ a distance $d_q$ between the neighbouring such lines of zeros mesured along the
exceptional SL (it is $L_k^l$ this time) is equal to $d_q=i\pi e^{-i\beta}/\sqrt{W_{n,\alpha}(\zeta_{l,q00}^{(k)})}$ and
vanishes with $q\to\infty$.

Fig.4a illustrates the exceptional lines for $\psi_{\alpha,\frac{n+2}{2}}(z)$ for even $n$ and
$\alpha=e^{-i\frac{\pi}{n}}$ occupied by zeros of this solution with the distinguished exceptional line
$L_\frac{n+2}{2}^{\frac{n}{4}-1}$ on which
the discribed above details of the zeros distribution are shown.

To formulate the corresponding theorem for the critical cases of SG's, i.e. for real $\lambda$,
we have to notice that $C_{cut}\setminus D_k=\p D_k\cup S_{-k}$ so that $\p D_k$ contains the point
$z_{-k}$ together with their three SL's which emerge from it. This is just a set $L_k^{-k}$ of exceptional SL's emerging
from $z_{-k}$.

Let us note further a role played in the critical cases of SG's by the integral
$\lambda\int_{z_k}^{z_{-k}}\sqrt{W_{n,\alpha}(y)}dy\equiv\lambda I_k$. For $\lambda$ real and positive it can always be
represented as $I_k\lambda=-(r+R)i\pi$ where
$r\geq 0$ is an integer and $|R|\leq\fr$. Both $r$ and $R$ depend, of course, on $\lambda$ and due this dependence
$R$ can take on an arbitrary value from its domain of variation. Conversely, fixing $R$ on some $R_0$ we choose
in this way an infinite sequence $\lambda_{k,r}(R_0)=-\frac{i\pi}{I_k}(r+R_0),\;r=0,1,2,...$, of $\lambda$'s growing to
infinity and such that
$R(\lambda_{k,r}(R_0))=R_0$. We shall maintain further for such sequences a description {\it regular} as well as for the limit
$\lambda\to\infty$ itself using these sequences if $|R|<\fr$ while we call them {\it singular} if $|R|=\fr$. As previously all
other sequences of $\lambda$'s not restricted by any condition will be again called {\it free}.

It should be stressed however that as it follows from the form of the $\chi$-factors of FS's and what will be seen later
when semiclassical limit $\lambda\to\infty$ of particular formulae will be taken that to apply regular limits one
needs to statisfy the inequality $\cos(R\pi)\gg \frac{\lambda_0}{\lambda}>0$. However this condition is always
satisfied for any $|R|<\fr$ when $\lambda$ is sufficiently large.

\begin{tabular}{c}
\psfig{figure=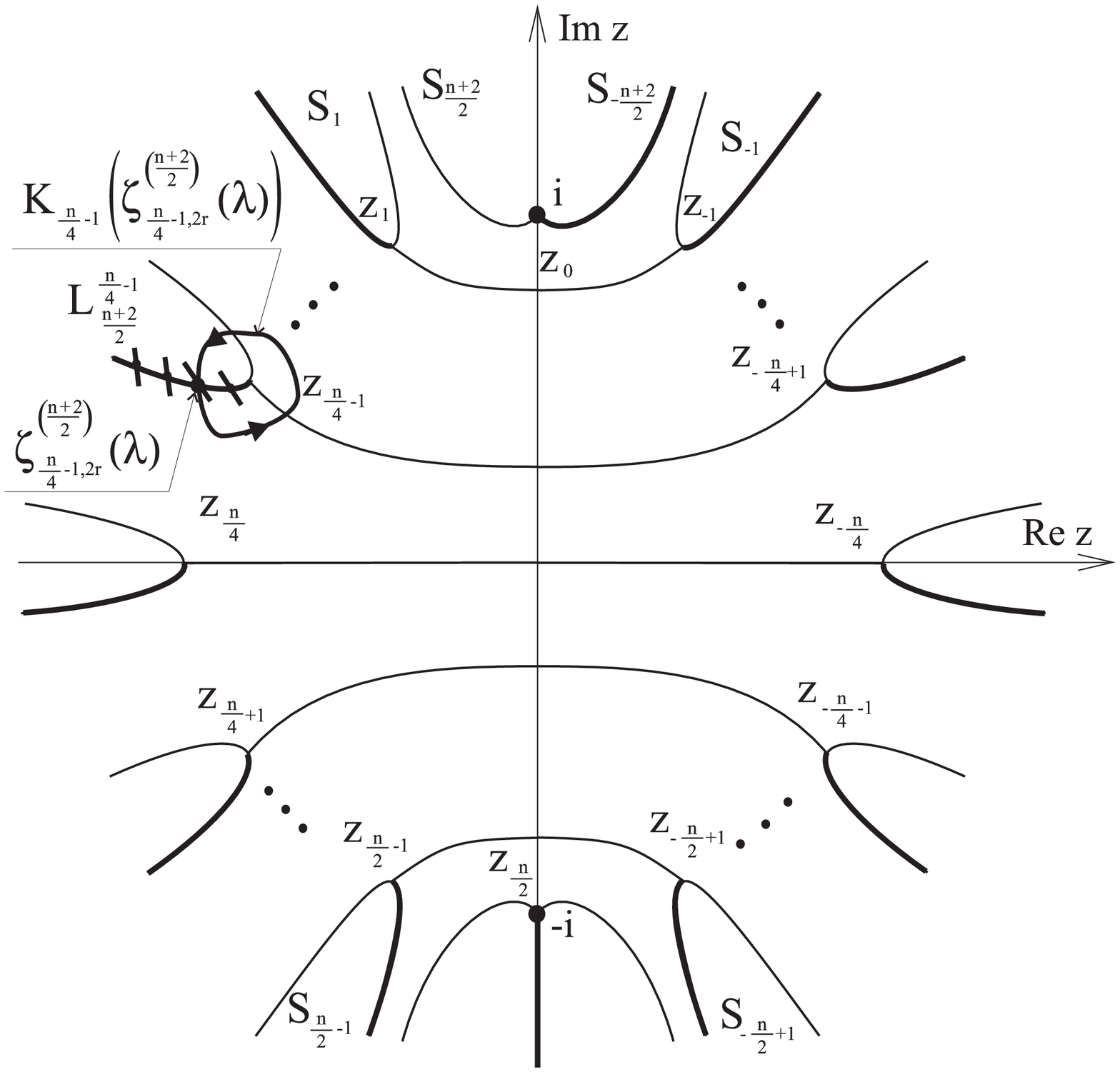,width=11cm}\\
Fig.4a SG for an even $n$ and $\alpha=1$. The bold lines are exceptional SL's along which\\
zeros of $\psi_{\frac{n+2}{2}}(z)$ are distributed in the regular limit $\lambda\to\infty$.\\
On one of them details of such ditribution are shown
\end{tabular}

\vskip 12pt

In {\bf Theorem 2a} we have used regular sequences of $\lambda$ conditioned by fixed value of $\Lambda$. In
{\bf Theorem 2b} below the regular limit $\lambda\to\infty$ will be also considered when $R$ has a fixed value. This is
essentially the main difference between the critical and non critical cases in taking the limit $\lambda\to\infty$.

{\bf Theorem 2b}

{\it Zeros $\zeta_{l,qr}^{(k)}(\lambda),\;|\lambda|=[|\lambda|]+\Lambda,\;l=0,\pm 1,\pm 2,...,l\neq -k,\;q=1,2,...,\;
r=0,\pm 1,\pm 2,...,$ of
$\psi_{\alpha,k}(z,|\lambda|)$, i.e. in the critical cases, in the regular limit $[|\lambda|]\to\infty$, i.e. with
fixed $\Lambda$, are distributed on $C_{cut}$ uniquely along the corresponding exceptional SL's according to the formulae}:
\be
\int_{K_l(\zeta_{l,qr}^{(k)}(\lambda))}\ll(\fr\sqrt{W_{n,\alpha}(y)}-
\frac{1}{2\lambda}Z_{\alpha,k}(y,\lambda)\r)dy=\pm\ll(q[|\lambda|]+r-\frac{1}{4}\r)\frac{i\pi}{\lambda}
\label{14b}
\ee
{\it where $K_l(\zeta_{l,qr}^{(k)}(\lambda))$ is a contour which starts and ends at
$\zeta_{l,qr}^{(k)}(\lambda)$ rounding the turning point $z_l$ anticlockwise.

Zeros $\zeta_{l,qr}^{(k)}(\lambda)$ have the following semiclassical expansion:}
\be
\zeta_{l,qr}^{(k)}(\lambda)=\sum_{p\geq 0}\frac{1}{\lambda^p}\zeta_{l,qrp}^{(k)}(\Lambda)
\label{143}
\ee
{\it with two first terms given by}:
\be
\int_{K_l(\zeta_{l,qr0}^{(k)})}\fr\sqrt{W_{n,\alpha}(y)}dy=
\int_{z_l}^{\zeta_{l,qr0}^{(k)}}\sqrt{W_{n,\alpha}(y)}dy=\pm qi\pi \nn\\
\zeta_{l,qr1}^{(k)}(\Lambda)=
\pm(r-q\Lambda-\frac{1}{4})\frac{i\pi }{\sqrt{W_{n,\alpha}(\zeta_{l,qr0}^{(k)})}}
\label{144}
\ee

{\it For $q=0$ we have instead $\zeta_{l,0r0}^{(k)}(\Lambda)\equiv z_l$ and}
\be
\int_{z_l}^{z_l+\zeta_{l,0r1}^{(k)}(\Lambda)/\lambda}\sqrt{W_{n,\alpha}(y)}dy=
(r-\frac{1}{4})\frac{i\pi}{\lambda}\nn\\
r>\frac{|\lambda|}{\pi}\limsup_{|\phi|\leq\pi}\ll|\int_{z_l}^{z_l+\epsilon e^{i\phi}}\sqrt{W_{n,\alpha}(y)}dy\r|
\label{144a}
\ee
{\it as well as}
\be
\zeta_{l,0r2}^{(k)}(\Lambda)=\frac{1}{8}\frac{\int_{K_l(z_l+\zeta_{l,0r1}^{(k)}(\Lambda)/\lambda)}X_{\alpha,1}(y)dy}
{\sqrt{W_{n,\alpha}(z_l+\zeta_{l,0r1}^{(k)}(\Lambda)/\lambda)}}
\label{144b}
\ee

{\it The signs $\pm$ above are to be chosen the same as of
$\Im\ll(\int_{z_{l}}^{z} \sqrt{W_{n,\alpha}(y)}dy\r)$ on the corresponding exceptional lines.

In the case $l=k$ the number
$q$ is bounded, i.e. $q\leq |I_k|/\pi$ and in} \mref{144} {\it the minus sign is to be chosen according to our earlier
conventions}.

{\it Additionally in a regular limit $\lambda_{k,s}\to\infty$, i.e. with $R$ fixed, where
$\lambda_{k,s}=-\frac{s+R}{I_k}i\pi=[\lambda_{k,s}]+\Lambda_{k,s}(R),\;s=0,1,2,...$, there are two infinite sequences
of zeros $\zeta_{-k,qr}^{(k)\pm},\;q=1,2,...,\;r=0,\pm 1,\pm 2,...,$, of
$\psi_{\alpha,k}(z,|\lambda|)$ distributed along the two infinite SL's of the sector $S_{-k}$ according to the
following rules}:
\be
\int_{z_{-k}}^{\zeta_{-k,qr}^{(k)\pm}}\sqrt{W_{n,\alpha}(y)}dy=-\ll(q[\lambda_{k,s}]+r-\frac{1}{4}+\frac{R}{2}\r)
\frac{i\pi}{\lambda_{k,s}}+
\frac{1}{4\lambda_{k,s}}\oint_{K_k}Z_{\alpha,k}dy\pm\nn\\
\frac{1}{2\lambda_{k,s}}\ln2\cos\ll(R\pi+\fr\Im\oint_{K_k}Z_{\alpha,k}dy\r)\pm
\frac{1}{2\lambda_{k,s}}\int_{K_{-k}(\zeta_{-k,qr}^{(k)\pm})}Z_{\alpha,k}dy
\label{14c}
\ee
{\it with the following first coefficients of the corresponding semiclassical expansion of $\zeta_{-k,qr}^{(k)\pm}$}:
\be
\int_{z_{-k}}^{\zeta_{-k,qr0}^{(k)\pm}}\sqrt{W_{n,\alpha}(y)}dy=-qi\pi\nn\\
\zeta_{-k,qr1}^{(k)\pm}(R)=-\ll(r-q\Lambda_{k,s}(R)-\frac{1}{4}+\frac{R}{2}
\mp\fr\ln2\cos(R\pi)\r)\frac{i\pi}{\sqrt{W_{n,\alpha}(\zeta_{-k,qr0}^{(k)\pm})}}
\label{145}
\ee
{\it where the plus sign corresponds to a vicinity of the SL being the upper boundary of $S_{-k}$ while the minus one
to a vicinity of its lower boundary}.

Fig.4b shows all the exceptional SL's of the odd $n$ case occupied by zeros of $\psi_{\alpha,k}(z,\lambda)$ in the
limits considered in the above theorem.

All the above three theorems have been proved in Appendix 2.

It follows therefore from the notes made above that the cases $R=\pm \frac{1}{2}$ have to be considered separately. It
should not be surprising that they are just the quantized cases of the solution $\psi_{\alpha,k}(z,\lambda)$ considered.

\section{High energy limit of loci of zeros of fundamental solutions for the potential $W_n(z,\lambda)+1$}

\hskip+2em Let us denote $-i\alpha z$ by $\zeta$ and $\lambda^{-\frac{2}{n+2}}$ by $\eta$  and let
$|\arg\lambda|<\frac{\pi}{n}$. Then for the polynomial $W_n(z,\lambda)$ we have:

\be
W_n(z,\lambda)\to {\tilde W}_n(\zeta,\eta)\equiv W_n(i\alpha^{-1}\zeta,\eta^{-\frac{n+2}{2}})
=\zeta^n-1+\sum_{k=1}^{n-1}b_{n-k}'\eta^k\zeta^{n-k}\nn\\
b_{n-k}'=b_{n-k}(-i\alpha)^{\frac{2k}{n+2}}
\label{15}
\ee

Let us note that bounding $|\arg\lambda|$ by $\frac{\pi}{n}$ we always get the standard SG's corresponding to $W_{n,\alpha}(z)$
as a limit of the corresponding SG's related to $W_n(z,\lambda)$ when $\lambda\to\infty$.

It follows from the above form of ${\tilde W}_n(\zeta,\eta)$ that if $\zeta_0$ is its root for $\eta=0$ then, by the
implicite function theorem, there is also such a root $\zeta(\eta)$ of ${\tilde W}_n(\zeta,\eta)$ which is close to
$\zeta_0$ for $\eta$ sufficiently close to zero, i.e. $\zeta(\eta)=\zeta_0-\frac{1}{n}b_{n-1}'\eta+O(\eta^2)$ for
$\eta\to 0$. It means of course
that for sufficiently large $\lambda$ all the roots of $W_n(z,\lambda)$ are simple and are close to the corresponding
roots of $W_{n,\alpha}(z)$, i.e to each root $z_k$ of the latter potential there is the root $z_k(\lambda)$ of the former one
such that $z_k(\lambda)=z_k-i\frac{1}{n\alpha}b_{n-1}'\lambda^{-\frac{2}{n+2}}+O(\lambda^{-\frac{4}{n+2}})$ for
$|\lambda|\to\infty,\;k=0,\pm 1,...$.

Let us re-examin for the potential considered the respective notions of SG's and FS's and remaining ones related to them.
A necessity to do it is caused by the $\lambda$-dependence of the
potential $W_n(z,\lambda)+1$ and as it is seen from \mref{15} this dependence is by powers of $\lambda^{-\frac{2}{n+2}}$.
Such a dependence complicates an analysis of $\lambda$-dependence of SG's, FS's and other notions. It is easier to make
the respective analyses by a substitution $\lambda\to\lambda^\frac{n+2}{2}$ in the corresponding Schr\"odinger equation
\mref{2} with the potential $W_n(z,\lambda)+1$ by which the potential itself convert to depend on natural powers of
$\lambda^{-1}$ and making simultanuously the same substitution for the SE with the potential $W_{n,\alpha}(z)+1$.

\vskip 12pt

\begin{tabular}{c}
\psfig{figure=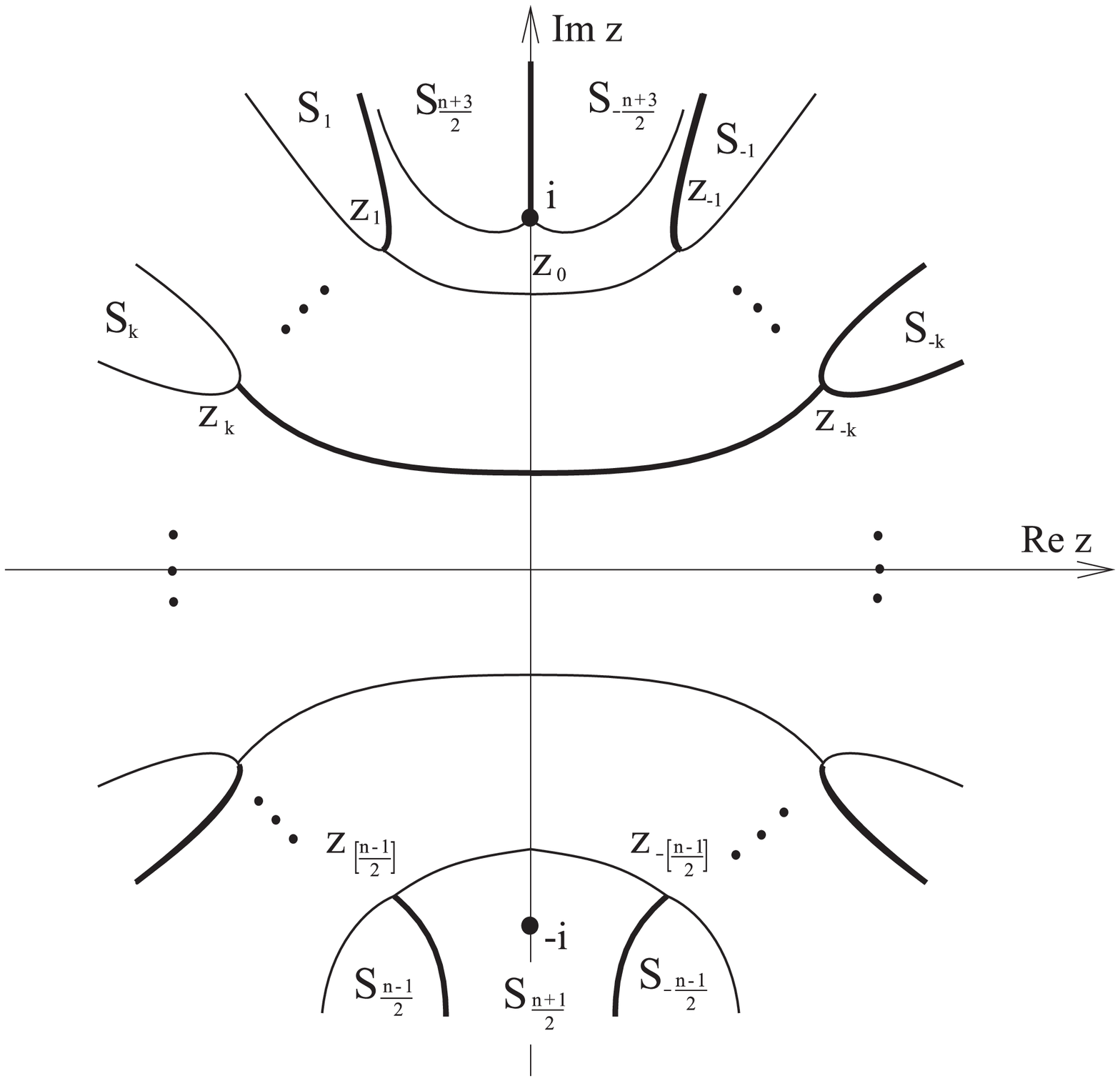,width=11cm}\\
Fig.4b SG for an odd $n$ and $\alpha=1$. Bold lines are exceptional SL's along which\\
zeros of $\psi_{\frac{n+2}{2}}(z)$ are distributed in the regular limit $\lambda\to\infty$
\end{tabular}

\vskip 12pt

The corresponding SG's can be now defined by using the new $\lambda$-parameter for both the potentials. It does not
disturb much properties of SG's for the potential $W_{n,\alpha}(z)+1$ except that it makes these graphs more sensitive
for $\lambda$-changes (because $\lambda$ is now powered by $(n+2)/2$), i.e. the graphs come back to their initial forms when
$\lambda$ is rotated by the angle $\phi=2\pi/(n+2)$ rather than by $\pi$ as previously, for example, and this is just this
new angle which is now a period for such graphs' changes.

In the case of the potential $W_n(z,\lambda)+1$ the corresponding graphs defined by \mref{4} where $\lambda$ is
substituted by $\lambda^\frac{n+2}{2}$ depend now on $\lambda$ and if $\lambda$ changes
it needs a total change equal to $2\pi$ of its argument to make the graphs coming back to their initial positions.
Changing absolute value of $\lambda$ alone also forces the graphs to change while the graphs corresponding to
$W_{n,\alpha}(z)+1$ are insensitive on such a change of $\lambda$.

All these do not prevent us however to define the
corresponding FS's by the formulae \mref{10}-\mref{13} together with the corresponding definitions of canonical
paths, canonical domains, etc, including also constructions of domains $D_{k,\epsilon}$ contained in canonical domains
$D_k$ together with $\epsilon$-vicinities  $V_{k,\epsilon}$ of their respective boundaries
$\partial D_k,\;k=0,\pm 1,...,$ and with {\it exceptional} SL's related to them. We have only to remember that
all the above notions depend now on $\lambda$ and can change with it continuously.

Nevertheless despite the mentioned $\lambda$-dependence {\bf Lemma} of sec.4 remains valid also in the case of the
potential $W_n(z,\lambda)+1$. This can be argued as follows.

Consider a fundamental solution $\psi_k(z,\lambda)$ to the SE \mref{2} corresponding to the potential $W_n(z,\lambda)+1$
and a respective domain $D_{k,\epsilon}(\lambda)$ where this solution can be continued canonically to any of its point
and let $|\lambda|>\lambda_0$ with $\lambda_0$ sufficiently large to find each $z_k(\lambda)$ inside corresponding circle
vicinities $\Delta_k(\epsilon),\;k=1,...n$.

By the same arguments and with obvious changes the inequality \mref{14} in the domain $D_{k,\epsilon}(\lambda)$ can be
written for the solution $\psi_k(z,\lambda)$ as follows:
\be
|\chi_k(z,\lambda)-1|\leq \exp\ll[\frac{C_{\epsilon}(\lambda)}{\lambda_0^\frac{n+2}{2}}\r]-1\nn\\
C_{\epsilon}(\lambda)=
\liminf_{\gamma_k(z),\;z\in C_{cut}(\epsilon),\;k=1,...,n+2}\int_{\gamma_k(z)}|\omega(\xi,\lambda)d\xi|<\infty
\label{16}
\ee

Therefore {\bf Lemma} of sec.4 sounds now (see App.2 for a proof):

{\bf Lemma'}

{\it In the domain $D_{k,\epsilon}(\lambda)$ the factor $\chi_k(z,\lambda)$ of the FS solution $\psi_k(z,\lambda)$
constructed for the
potential $W_n(z,\lambda)+1$ satisfies the following bound}
\be
|\chi_k(z,\lambda)-1|\leq \exp\ll[\frac{C_{\epsilon}}{\lambda_0^\frac{n+2}{2}(\epsilon)}\r]-1,
\;\;\;\;\;\;|\lambda|>\lambda_0(\epsilon)\nn\\
C_{\epsilon}=C_{\alpha;\epsilon}+\epsilon
\label{161}
\ee
{\it with $\lambda_0(\epsilon)$ sufficiently large and with fixed but arbitrary small $\epsilon$}

Therefore {\bf Theorem 1} remains valid also for the potential $W_n(z,\lambda)+1$ with no changes.

To formulate analogues of {\bf Theorem 2a} and {\bf 2b} we have to establish the corresponding forms of the semiclassical
expansions \mref{551}-\mref{553} for the case of the potential $W_n(z,\lambda)+1$. These expansions have to be a little bit
different because of a dependence of the potential on $\lambda$.

First we note that the $\chi$-factors of the FS's $\psi_k(z,\lambda)$ with the potential $W_n(z,\lambda)+1$ and $\lambda$ changed
as mentioned earlier satisfy the following equation:
\be
W_n^{-\frac{1}{4}}(z,\lambda)\ll(W_n^{-\frac{1}{4}}(z,\lambda)\chi_k(z,\lambda)\r)''+
2\sigma_k\lambda^{\frac{n+2}{2}}\chi_k'(z,\lambda)=0
\label{162}
\ee
which can be transformed to the (pseudo-)integral equation of the form
\be
\chi_k(z,\lambda)=1-\frac{\sigma_k}{2\lambda^{\frac{n+2}{2}}}
\int_{\infty_k}^zW_n^{-\frac{1}{4}}(y,\lambda)\ll(W_n^{-\frac{1}{4}}(y,\lambda)\chi_k(z,\lambda)\r)''dy
\label{163}
\ee
corresponding to the $\chi_k$-factor of the FS defined in the sector $S_k(\lambda)$.

An attempt to get a solution for $\chi_k$ from \mref{163} by iterations leads to a divergent series which however appears
to be just a semiclassical expansion for $\chi_k(z,\lambda)$ of the form:
\be
\chi_k(z,\lambda)\sim\chi_k^{as}(z,\lambda)=
\sum_{p{\geq}0}\left(-\frac{\sigma_k}{2\lambda^\frac{n+2}{2}}\right)^p{\tilde Y}_{k,p}(z,\lambda)
\label{164}
\ee
where ${\tilde Y}_{k,p}(z,\lambda)$ are given by \mref{552} with $W_{n,\alpha}(z)$ substituted by $W_n(z,\lambda)$.

If further using \mref{552} (with the substitutions mentioned) we make the following expansion:
\be
{\tilde Y}_{k,p}(z,\lambda)=\sum_{q\geq 0}{\tilde Y}_{k;p,q}(z)\lambda^{-q}\nn\\
{\tilde Y}_{k;0,q}(z)\equiv\delta_{0q}
\label{165}
\ee
then \mref{164} can be given the following forms depending on a parity of $n$:
\be
\chi_k^{as}(z,\lambda)=\ll\{\begin{array}{lr}
\sum_{p{\geq}0}\sum_{q=0}^m\frac{Y_{k;pq}(z)}{\lambda^{(m+1)p+q}},&n=2m\\
&\\
\sum_{p{\geq}0}\frac{1}{2^{2p}}\sum_{q=0}^{2m+2}\frac{Y_{k;2p,q}(z)}{\lambda^{(2m+3)p+q}}-&\\
\frac{\sigma_k}{2\lambda^{\frac{2m+3}{2}}}
\sum_{p{\geq}0}\frac{1}{2^{2p}}\sum_{q=0}^{2m+2}\frac{Y_{k;2p+1,q}(z)}{\lambda^{(2m+3)p+q}},&
n=2m+1
\end{array}\r.
\label{166}
\ee
where the coefficients of the last expansions are given by the ones of the expansion \mref{165} as:
\be
\begin{array}{lr}
Y_{k;pq}(z)=\sum_{r=0}^p\ll(-\frac{\sigma_k}{2}\r)^r{\tilde Y}_{k;r,(m+1)(p-r)+q}(z),&n=2m\\
\ll.\begin{array}{lr}
Y_{k;2p,q}(z)=\sum_{r=0}^p2^{2(p-r)}{\tilde Y}_{k;2r,(2m+3)(p-r)+q}(z)&\\
Y_{k;2p+1,q}(z)=\sum_{r=0}^p2^{2(p-r)}{\tilde Y}_{k;2r+1,(2m+3)(p-r)+q}(z)&
\end{array}\r\}&n=2m+1
\end{array}
\label{167}
\ee

The exponential representations of $\chi_k^{as}(z,\lambda)$ have similar forms to \mref{551}, i.e.
\be
\chi_k^{as}(z,\lambda)=\exp\ll(\int_{\infty_k}^zZ_k(y,\lambda)dy\r)\nn\\
Z_k(z,\lambda)=\ll\{
\begin{array}{lr}
\sum_{p{\geq}1}\sum_{q=0}^m\frac{X_{k;pq}(z)}{\lambda^{(m+1)p+q}},&n=2m\\
&\\
\sum_{p{\geq}1}\frac{1}{2^{2p}}\sum_{q=0}^{2m+2}\frac{X_{k;2p,q}(z)}{\lambda^{(2m+3)p+q}}-&\\
\frac{\sigma_k}{2\lambda^{\frac{2m+3}{2}}}\sum_{p{\geq}0}\frac{1}{2^{2p}}\sum_{q=0}^{2m+2}\frac{X_{k;2p+1,q}(z)}{\lambda^{(2m+3)p+q}},&
n=2m+1
\end{array}\r\}=\nn\\
\frac{1}{\chi_k^{as}(z,\lambda)}\frac{d\chi_k^{as}(z,\lambda)}{dz}
\label{168}
\ee

However the corresponding recurrent relations for $X_{k;pq}(z)$ are now much more complicated. Nevertheless since
$Z_k(z,\lambda)$ satisfy the equations:
\be
W_n^{-\frac{1}{4}}\ll(W_n^{-\frac{1}{4}}\r)''+
2\ll(W_n^{-\frac{1}{4}}\ll(W_n^{-\frac{1}{4}}\r)'+\sigma_k\lambda^{\frac{n+2}{2}}\r)Z_k+
W_n^{-\frac{1}{2}}\ll(Z_k^2+Z_k'\r)=0
\label{170}
\ee
then the partition
$Z_k(z,\lambda)=Z^+(z,\lambda)+\sigma_kZ^-(z,\lambda)$ still can be done with the same properties \mref{557} for
$Z^+(z,\lambda)$ and the corresponding conlusions about its analytical properties on the $z$-plane.

Unlike the case of the potential $W_{n,\alpha}(z)+1$ we have to use also the following asymptotic expansion of
$\sqrt{W_n(z,\lambda)}$:
\be
\sqrt{W_n(z,\lambda)}=\sqrt{W_{n,\alpha}(z)}+\sum_{p{\geq}1}\frac{W_{n,p}(z)}{\lambda^p}
\label{169}
\ee
as well as the asymptotic expansions of the limit loci of zeros $\zeta_l^{(k)}(\lambda),\;l=0,\pm 1,\pm 2,...,$ of the
FS $\psi_k(z,\lambda)$:
\be
\zeta_l^{(k)}(\lambda)=
\ll\{\begin{array}{lr}
\sum_{p{\geq}0}\sum_{q=0}^m\frac{\zeta_{l;p,q}^{(k)}}{\lambda^{(m+1)p+q}}&n=2m\\
\sum_{p{\geq}0}\sum_{q=0}^{2m+2}\frac{\zeta_{l;2p,q}^{(k)}}{\lambda^{(2m+3)p+q}}+&\\
\frac{1}{\lambda^{\frac{2m+3}{2}}}
\sum_{p{\geq}0}\sum_{q=0}^{2m+2}\frac{\zeta_{l;2p+1,q}^{(k)}}{\lambda^{(2m+3)p+q}},&n=2m+1
\end{array}\r.
\label{170}
\ee

Having done properly the semiclassical expansions of the respective quantities in order to be as close as possible to the
previous formulations of {\bf Theorem 2a} and {\bf 2b}, we can come back to the previous form of $\lambda$-dependence
by a substitution in all the above formulae $\lambda^\frac{n+2}{2}$ back by $\lambda$ itself. Then we can formulate
the following theorems analogous to {\bf Theorem 2a} and {\bf 2b}.

{\bf Theorem 3a}

{\it In the non-critical case and in the regular limit $\lambda\to\infty$ zeros
$\zeta_{l,qr}^{(k)}(\lambda),\;|\lambda|=[|\lambda|]+\Lambda,\;l=0,\pm 1,\pm 2,...,\;q=1,2,...,\;
r=0,\pm 1,\pm 2,...,$ of
$\psi_k(z,|\lambda|e^{i\beta}),\;0<|\beta|<\frac{\pi}{n}$ are
distributed on $C_{cut}$ uniquely along the corresponding exceptional SL's according to the formulae}:
\be
\int_{K_l(\zeta_{l,qr}^{(k)}(\lambda))}\ll(\fr\sqrt{W_n(y,\lambda)}-
\frac{1}{2\lambda}Z_k(y,\lambda)\r)dy=\pm\ll(q[|\lambda|]+r-\frac{1}{4}\r)\frac{i\pi}{\lambda}
\label{17a}
\ee
{\it where $K_l(\zeta_{l,qr}^{(k)}(\lambda))$ is a contour which starts and ends at
$\zeta_{l,qr}^{(k)}(\lambda)$ rounding the turning point $z_l$ anticlockwise.

The two lowest terms of the semiclassical expansions of zeros $\zeta_{l,qr}^{(k)}(\lambda)$ given by} \mref{170}
{\it are the following}:
\be
\int_{K_l(\zeta_{l,qr;0,0}^{(k)})}\fr\sqrt{W_{n,\alpha}(y)}dy=
\int_{z_l}^{\zeta_{l,qr;0,0}^{(k)}}\sqrt{W_{n,\alpha}(y)}dy=\pm qi\pi e^{-i\beta}\nn\\
\zeta_{l,qr;0,1}^{(k)}(\Lambda)=
\pm(r-q\Lambda-\frac{1}{4})\frac{i\pi e^{-i\beta}}{\sqrt{W_{n,\alpha}(\zeta_{l,qr;0,0}^{(k)})}}
\label{172}
\ee
{\it and are calculated from} \mref{17a} {\it according to the formula} \mref{A234} {\it of} App.2.

{\it The signs $\pm$ above are to be chosen to agree with signs of
$\Im\ll(e^{i\beta}\int_{z_{l}}^{z} \sqrt{W_n(y,\lambda)}dy\r)$ on the corresponding exceptional lines}.

Suppose now that there is an inner SL linking the roots $z_{k_0},z_{-k_0}$ of $W_n(z,\lambda)$ while the others are
absent (see Fig.5b). In the limit case $\lambda\to\infty$ it is possible only for $\arg\lambda\neq 0$ but with
$\arg\lambda\sim|\lambda|^{-\frac{2}{n+2}}$ (see App.3). Then for the assumed arrangement of SL's we have the following:

{\bf Theorem 3b}

{\it In the critical case, when there is inner SL linking $z_{k_0}$ with $z_{-k_0}$, and in the regular limit
$\lambda\to\infty$ zeros $\zeta_{l,qr}^{(k)}(\lambda),\;|\lambda|=[|\lambda|]+\Lambda,\;l=0,\pm 1,\pm 2,...,
\;q=1,2,...,\;r=0,\pm 1,\pm 2,...,$ of
$\psi_k(z,|\lambda|e^{i\beta}),\;0<|\beta|<\frac{\pi}{n},\;k\neq k_0,-k_0$, are
distributed on $C_{cut}$ uniquely along the corresponding exceptional SL's according to the formulae}:
\be
\int_{K_l(\zeta_{l,qr}^{(k)}(\lambda))}\ll(\fr\sqrt{W_n(y,\lambda)}-
\frac{1}{2\lambda}Z_k(y,\lambda)\r)dy=\pm\ll(q[|\lambda|]+r-\frac{1}{4}\r)\frac{i\pi}{\lambda}
\label{17a}
\ee
{\it where $K_l(\zeta_{l,qr}^{(k)}(\lambda))$ is a contour which starts and ends at
$\zeta_{l,qr}^{(k)}(\lambda)$ rounding the turning point $z_l$ anticlockwise.}

\begin{tabular}{c}
\psfig{figure=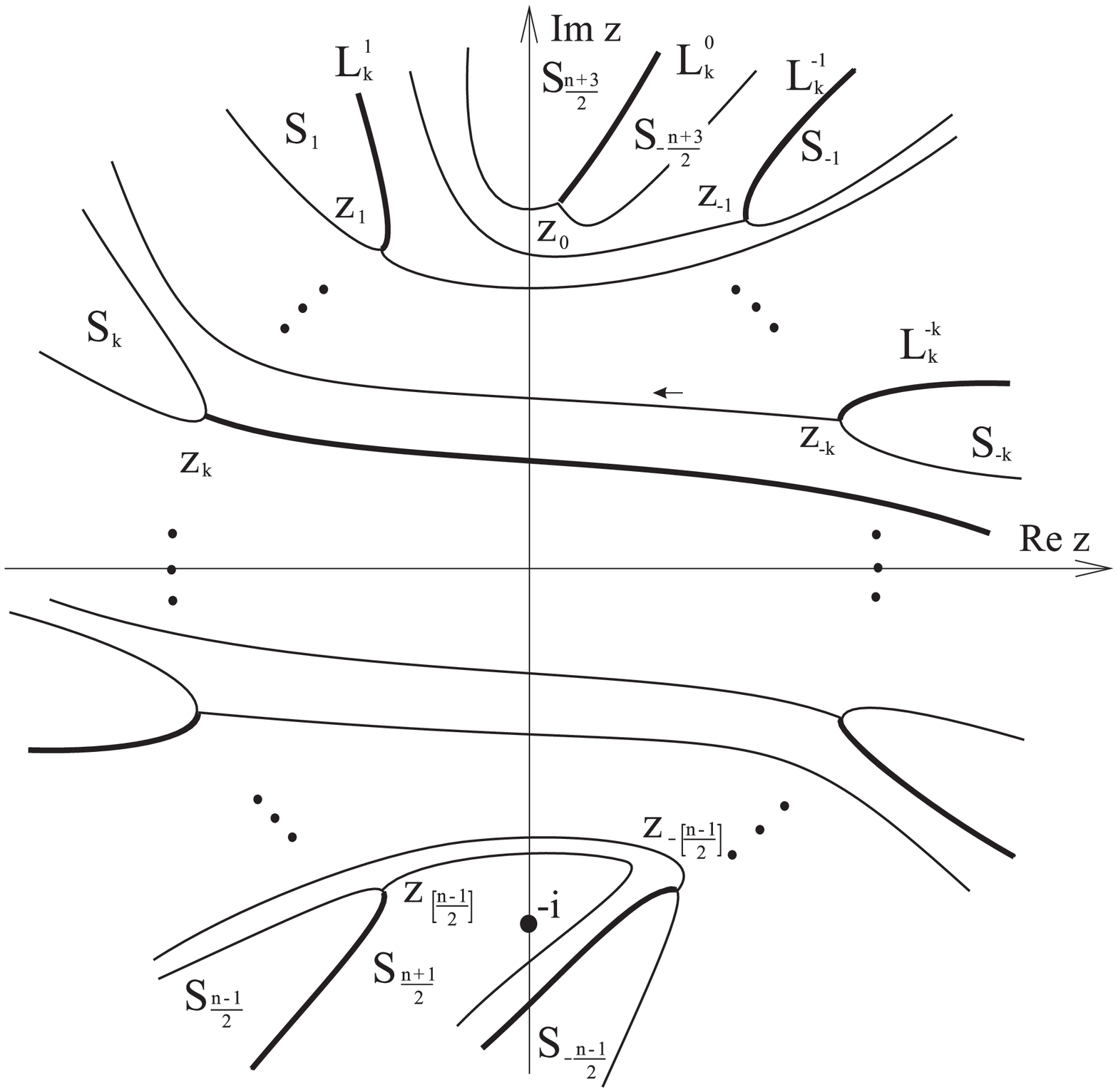,width=11cm}\\
Fig.5a The regular limit $\lambda\to\infty$ of zeros of $\psi_k(z)$ for the potential\\
$W_n(z)+1$ ($n$ - odd). The non-critical case. The bold lines are exceptional SL's
\end{tabular}

\vskip 12pt

{\it The two lowest terms of the semiclassical expansions of zeros $\zeta_{l,qr}^{(k)}(\lambda)$ given by} \mref{170}
{\it are the following}:
\be
\int_{K_l(\zeta_{l,qr;0,0}^{(k)})}\fr\sqrt{W_{n,\alpha}(y)}dy=
\int_{z_l}^{\zeta_{l,qr;0,0}^{(k)}}\sqrt{W_{n,\alpha}(y)}dy=\pm qi\pi e^{-i\beta}\nn\\
\zeta_{l,qr;0,1}^{(k)}(\Lambda)=
\pm(r-q\Lambda-\frac{1}{4})\frac{i\pi e^{-i\beta}}{\sqrt{W_{n,\alpha}(\zeta_{l,qr;0,0}^{(k)})}}
\label{172}
\ee
{\it and are calculated from} \mref{17a} {\it according to the formula} \mref{A234} {\it of} App.2.

{\it The signs $\pm$ above are to be chosen to agree with signs of
$\Im\ll(e^{i\beta}\int_{z_{l}}^{z} \sqrt{W_n(y,\lambda)}dy\r)$ on the corresponding exceptional lines.

In the cases $l=k_0,-k_0$ the numbers
$q$ are bounded, i.e. $q\leq |I_{k_0}|/\pi$ with the minus sign in} \mref{172} {\it chosen by assumption}.

{\it Additionally in a regular limit $\lambda_{k_0,s}\to\infty$, i.e. with $R$ fixed, where
$\lambda_{k_0,s}=-\frac{s+R}{I_{k_0}}i\pi=[\lambda_{k_0,s}]+\Lambda_{k_0,s}(R),\;s=0,1,2,...$, there are two infinite sequences
of zeros $\zeta_{-k_0,qr}^{(k_0)\pm},\;q=1,2,...,\;r=0,\pm 1,\pm 2,...,$, of
$\psi_{k_0}(z,\lambda)$ distributed along the two SL's of the sector $S_{-k_0}$ according to the
following rules}:
\be
\int_{z_{-k_0}}^{\zeta_{-k_0,qr}^{(k_0)\pm}}\sqrt{W_n(y,\lambda)}dy=-\ll(q[\lambda_{k_0,s}]+r-\frac{1}{4}+\frac{R}{2}\r)
\frac{i\pi}{\lambda_{k_0,s}}+
\frac{1}{4\lambda_{k_0,s}}\oint_{K_{k_0}}Z_{k_0}dy\pm\nn\\
\frac{1}{2\lambda_{k_0,s}}\ln2\cos\ll(R\pi+\fr\Im\oint_{K_{k_0}}Z_{k_0}dy\r)\pm
\frac{1}{2\lambda_{k_0,s}}\int_{K_{-k_0}}(\zeta_{-k_0,qr}^{(k_0)\pm})Z_{k_0}dy
\label{173}
\ee
{\it with the following lowest coefficients of the corresponding semiclassical expansion of $\zeta_{-k_0,qr}^{(k_0)\pm}$}:
\be
\int_{z_{-k_0}}^{\zeta_{-k_0,qr;0,0}^{(k_0)\pm}}\sqrt{W_{n,\alpha}(y)}dy=-qi\pi\nn\\
\zeta_{-k_0,qr;0,1}^{(k_0)\pm}(R)=-\ll(r-q\Lambda_{k_0,s}(R)-\frac{1}{4}+\frac{R}{2}
\mp\fr\ln2\cos(R\pi)\r)\frac{i\pi}{\sqrt{W_{n,\alpha}(\zeta_{-k_0,qr;0,0}^{(k_0)\pm})}}
\label{174}
\ee
{\it where the plus sign corresponds to the SL being the upper boundary of $S_{-k_0}$ while the minus one
to its lower boundary.

Obviously the second part of this theorem applies also to the FS $\psi_{-k_0}(z,\lambda)$ with appropriate changes.}

\section{High energy behaviour of zeros for quantized energy}

\hskip+2em By a quantization of energy we mean an identification of some two chosen fundamental solutions. However such a
choice is in fact quite limited particularly when the high energy limit, $|\lambda|\to\infty$, is taken.

First, in general, one can not identify two neighboured FS's since they are always linearly independent.

Secondly identifying in the high energy limit case any two not neighboured FS's we demand, as we will see below, the
sectors the fundamental solutions are defined in to be connected by an inner SL. This means that we can consider only the
standard SG cases, i.e. when these inner SL's are parallel to the real axis of the $C_{cut}$-plane. Therefore we can put
the energy quantization conditions only on the pairs $\psi_{\alpha,k}(z,\lambda),\psi_{\alpha,-k}(z,\lambda),
\;k=\pm 1,\pm 2,...$, for the potential $W_{n,\alpha}(z)$ and for analogous pairs for the potential $W_n(z,\lambda)+1$.
But as it follows from each of the formulae \mref{A34} and \mref{A35} of Appendix 2 the corresponding quantization
conditions for the potential $W_{n,\alpha}(z)$ when matching the pair
$\psi_{\alpha,k}(z,\lambda),\psi_{\alpha,-k}(z,\lambda)$ read:

\be
1+\exp\ll[2\int_{z_k}^{z_{-k}}\ll(\sigma_k\lambda\sqrt{W_{n,\alpha}(y)}+Z_k(y,\lambda)\r)dy\r]=0
\label{18}
\ee
or
\be
\int_{z_k}^{z_{-k}}\ll(\sigma_k\lambda_s\sqrt{W_{n,\alpha}(y)}+Z_k(y,\lambda_s)\r)dy=-(s+\fr)\pi i,\;\;\;\;s=0,1,2,...
\label{19}
\ee
what proves that in the limit $\lambda_s\to\infty$ the SL emerging from $z_k$ ends at $z_{-k}$, i.e. this is the inner SL.

\begin{tabular}{c}
\psfig{figure=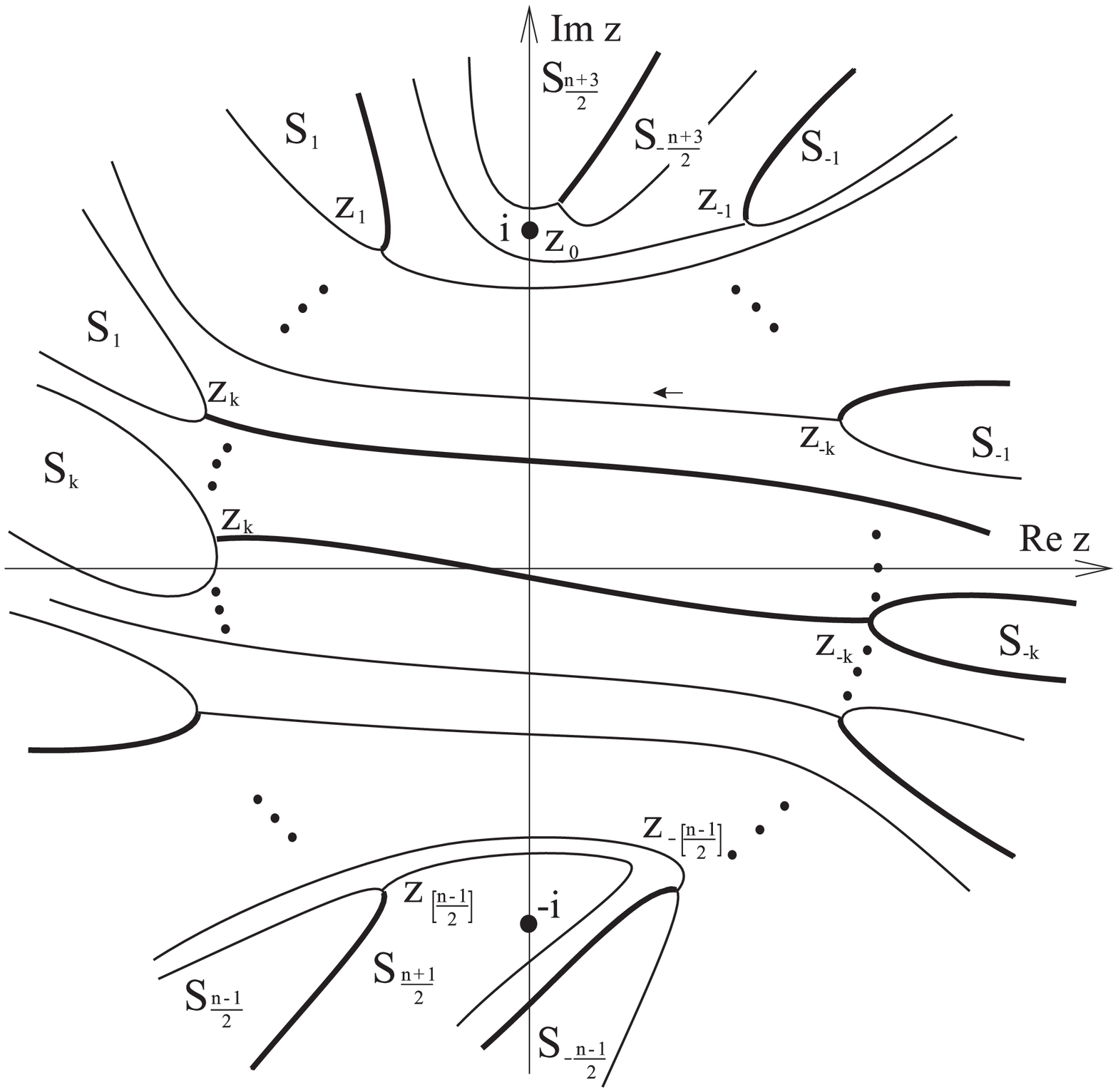,width=11cm}\\
Fig.5b The regular limit $\lambda\to\infty$ of zeros of $\psi_k(z)$ for the potential\\
$W_n(z)+1$ ($n$ - odd). The critical case. The bold lines are exceptional SL's
\end{tabular}

\vskip 12pt

The eq.\mref{19} thus proves that for the potential $W_{n,\alpha}(z)$ when the energy is quantized
there are inner SL's for each $k=\pm 1,\pm 2,...,$
but only for one value of $k$, i.e. this which satisfies \mref{19}, the energy is quantized and the corresponding FS's
coincide, i.e. $\psi_{\alpha,k}(z,\lambda_s)=i(-1)^s\psi_{\alpha,-k}(z,\lambda_s)$ as it follows from \mref{A34} or \mref{A35}.
However in the case of the potential
$W_n(z,\lambda)+1$ the SL satisfying eq.\mref{19} can be the unique inner one for the corresponding SG while the
remaining SL's of this graph emerge to infinity.

Let now eq.\mref{19} be satisfied for $k=k_0$. Then compairing \mref{19} with eq.\mref{14c} we see that
$R=\fr$ for this $k_0$ and therefore we can not apply the result given by \mref{14c} to this case. Nevertheless we can use the
fact that now $\psi_{\alpha,-k_0}(z,\lambda)$ coincides with $\psi_{\alpha,k_0}(z,\lambda)$ up to a constant and since
it can not vanish in its sector $S_{-k_0}$ so does $\psi_{\alpha,k_0}(z,\lambda)$, i.e. there are no longer roots of
$\psi_{\alpha,k_0}(z,\lambda)$ along the boundaries of the sector $S_{-k_0}$.

Nevertheless there are still such roots of $\psi_{\alpha,k}(z,\lambda)$ in the corresponding sectors
$S_{-k}$ for $k\neq k_0$.

Let us call the fundamental solution $\psi_{\alpha,k_0}(z,\lambda_s)$ {\it quantized} if the quantization conditions \mref{19}
is satisfied just for the number $k_0$. Of course it is satisfied also for the number $-k_0$ so the solution
$\psi_{\alpha,-k_0}(z,\lambda_s)$ is also quantized and both the solutions coincide up to a constant.

Therefore if we identify the quantized solutions $\psi_{\alpha,k_0}(z,\lambda_s)$ and $\psi_{\alpha,-k_0}(z,\lambda_s)$ then
an exceptional set for them is just a section $\bigcup_{r=1}^{n}L_{k_0}^r\cap\bigcup_{r=1}^{n}L_{-k_0}^r$ while for the
remaining FS's their exceptional sets are kept unchanged.

We thus come up to the following conclusion:

\begin{tabular}{c}
\psfig{figure=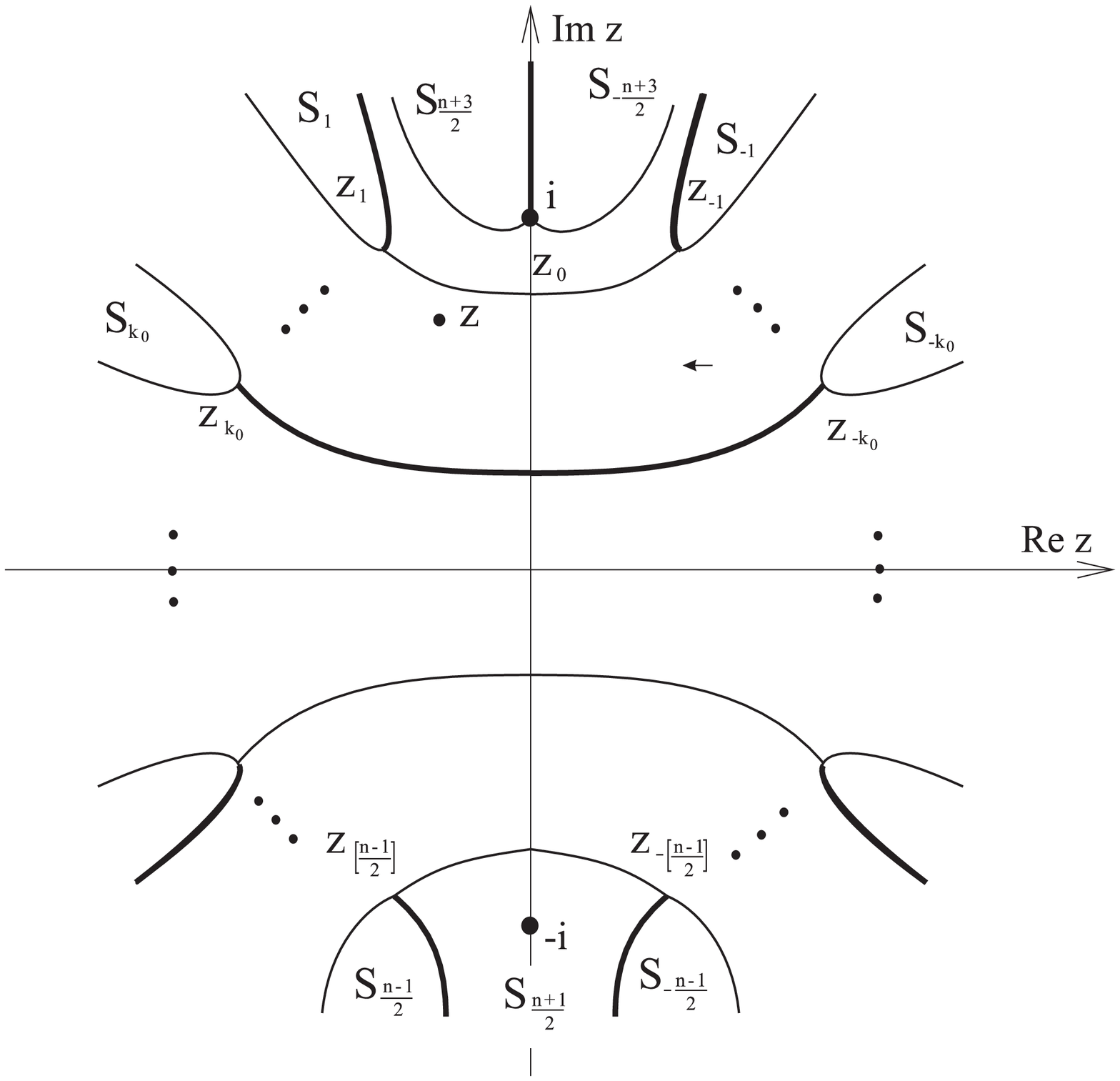,width=11cm}\\
Fig.5c The singular ($R=\fr$) limit $\lambda_r\to\infty$ of zeros of $\psi_{k_0}(z)$ for the potential\\
$W_{n,\alpha}(z)$ ($n$ - odd). The critical case. The bold lines are exceptional SL's
\end{tabular}

\vskip 12pt

{\bf Corollary 1a}

{\it In the case of the potential $W_{n,\alpha}(z)$ when the singular high energy limit $\lambda_s\to\infty$ (or
equivalently $s\to\infty$) are considered
roots of both the quantized FS's and the not quantized ones are distributed on the $C_{cut}$-plane
uniquely on the exceptional SL's corresponding to these solutions and the corresponding formulae of
{\bf Theorem 2b}, where $\lambda$ and $\Lambda$ should be substituted by $\lambda_s$ and $\Lambda_s$ respectievly, are
valid for these distributions excluding the formulae} \mref{14c} {\it and} \mref{145} {\it which
are nor longer valid for the quantized solutions} (see Fig.5c).

Exactly the same notes as above can be done with respect to the quantized solutions $\psi_{k_0}(z,\lambda_s)=
i(-1)^s\psi_{-k_0}(z,\lambda_s)$ corresponding to the potential $W_n(z,\lambda)+1$. They are now both deprived of zeros
lying on the infinite SL's emerging from the turning points $z_{k_0}(\lambda)$ and $z_{-k_0}(\lambda)$ while keeping their
zeros distributed on the inner SL linking these two turning points and on their remaining exceptional SL's.

It is therefore clear how the corresponding conclusion for the quantized high energy limit for the potential $W_n(z,\lambda)+1$
should sound.

{\bf Corollary 1b}

{\it In the singular high energy limit $\lambda_s\to\infty$ roots of FS's for the potential $W_n(z,\lambda)+1$ are distributed
uniquely on exceptional lines corresponding to these solutions according to the formulae of {\bf Theorems 3b},
where $\lambda$ and $\Lambda$ should be substituted by $\lambda_s$ and $\Lambda_s$ respectievly, except
the quantized solutions $\psi_{k_0}(z,\lambda_s)=
i(-1)^s\psi_{-k_0}(z,\lambda_s)$ for which the formulae} \mref{173} {\it and} \mref{174} {\it are nor longer valid}
(see Fig.5d).

\begin{tabular}{c}
\psfig{figure=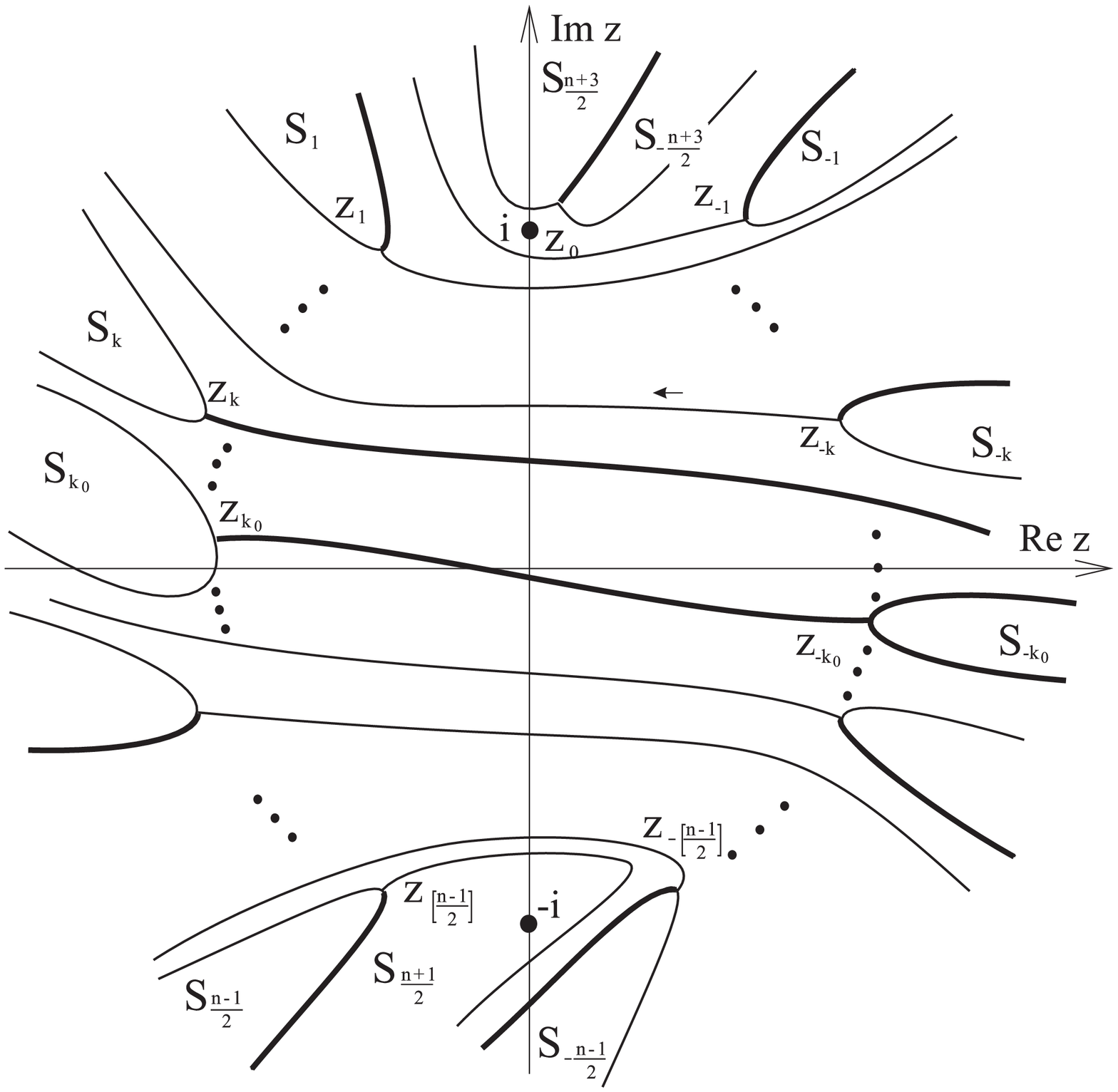,width=11cm}\\
Fig.5d The singular ($R=\fr$) limit $\lambda_r\to\infty$ of zeros of $\psi_{k_0}(z)$ for the potential\\
$W_n(z)+1$ ($n$ - odd). The critical case. The bold lines are exceptional SL's
\end{tabular}

\vskip 12pt

\section{Summary and discussion}

\hskip+2em In this paper we have shown that the high energy limit distributions of zeros of appropriately scaled
fundamental solutions for polynomial potentials can be described completely both for the quantized and non-quantized
cases of these solutions. The quantized cases were considered earlier by Eremenko {\it et al} \cite{8} where the authors pointed out that the
ESL's are the loci of zeros of FS's. We have completed their observation in this case giving a detailed description of
positions of these zeros on the ESL's. However we have considered the zeros distribution problem of FS's in general
showing that loci of zeros of FS's on their ESL's is their common property which has to be completed by {\bf Theorems
2b-3b} which find additional zeros of FS's outside the ESL's for the crititical cases of SG's.

However, to get stable patterns for zeros loci distributions of FS's we have been forced to consider regular asymptotic high
energy limit of these loci getting as a result island pictures for these distributions with the islands numbered
by $q$ in the corresponding theorems. Different regular limits (controlled by the $\Lambda$ and $R$ parameters) have
lead do different distributions of zeros inside each island with the latter distributions being controlled by the
$r$-parameter in {\bf Theorems 2-3}.

It is worth to note also that taking these regular limits stabilizes the limit zeros
distributions of FS's in the same way as the rescaling $z$-variable by energy $E$ in the initial polynomial potential
$P_n(z)$ stabilizes its zeros distribution in the limit $E\to\infty$ reducing it to the zeros loci of the polynomial
$(-i\alpha z)^n-1$.

\section*{Acknowledgment}

\hskip+2em I would like to thank to Boris Shapiro for focusing my attention on the problem of zeros of FS's as well as
for many discussions on it during my visit in the Department of Mathematics of Stockholm University.

\section*{Appendix 1}

\hskip+2em We shall show here that the integrals \mref{5} are all pure imaginary. We simply calculate them. To this
goal it is necessary to consider
$W_{n,\alpha}(z)$ on the cut $z$-plane. We make all these cuts from each turning point $z_k$ parallelly to the
real axis and running to $\Re z=-\infty$ if $z_k$ lies to the left from the imaginary axis or is placed on it and
running to $\Re z=+\infty$
in the opposite case. Then argument of $z-z_k$ will be taken from the interval
$(-\pi,\pi)$ for the first group of turning points taking the values $-\pi,\pi$  below and above the corresponding cuts
respectively (see Fig.1a). For the secod group of $z_k$ their arguments are taken from the interval $(0,2\pi)$ with the
values $0,2\pi$ above and below the cuts respectively.

Now, making in the integral \mref{5} the following change of variable $z\to\xi=(-i\alpha z)^n$ we are to calculate the
integrals
\be
\int_{z_{k}}^{z_{-k}} \sqrt{W_{n,\alpha}(z)}dz=\int_{C_k}\sqrt{W_{n,\alpha}(z)}dz=
-\frac{1}{i\alpha n}\int_{{\tilde C}_k}(\xi-1)^\fr\xi^\frac{1-n}{n}d\xi
\label{A11}
\ee
where the integration contour $C_k$ is an arc of a unit radius starting from the point $z_k$ and ending at $z=z_{-k}$,
running clockwise and avoiding all the met turning ponts $z_{k-1},...,z_{-k+1}$ (see Fig.1a) while ${\tilde C}_k$ is an
image of $C_k$ on the $\xi$-Riemann surface defined by the above change of variable (see Fig.6).

\begin{tabular}{c}
\psfig{figure=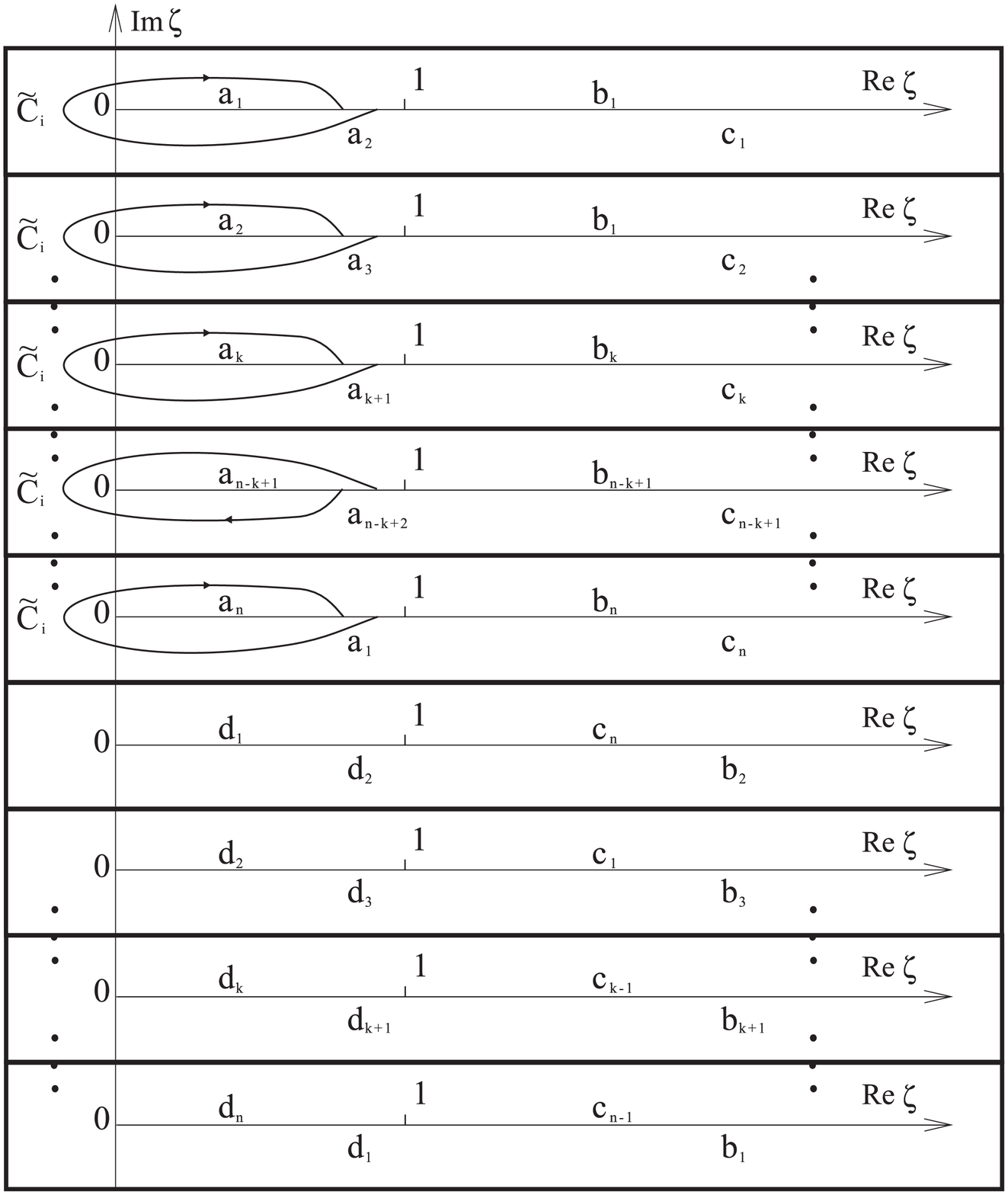,width=11cm}\\
Fig.6 $2n$ cut $\xi$-planes of Riemann surface of $(\xi-1)^\fr\xi^\frac{1-n}{n}$.
The letters $a_1,...,d_n$\\denote boundaries of cuts. Boundaries with the same letters are glued
\end{tabular}

\vskip 12pt

It is easy to note that
${\tilde C}_k$ is a collection of $2k$ or $2k-1$ unit radius circles on this $\xi$-Riemann surface depending on $\alpha$. The first
value corresponds to $\alpha =1$ and the second to $\alpha =e^{-\frac{i\pi}{n}}$. The first circle starts at the point
$\xi=1$ on the sheet on which $\arg\xi=2k\pi$. The latter argument is just the one of $\xi_k=(-i\alpha z_k)^n$. In Fig.6
this is the $k$-th sheet from above with $a_{k+1}$ as the lower boundary of the cut between $0$ and $1$ on this sheet.
Each next
circle is an image of successive arcs of the contour $C_k$. The last circle corresponds to the unit radius arc between
the points $z_{-k+1}$ and $z_{-k}$ and it ends at unity on the $n-k+1$-th sheet.

It is easy to note that
${\tilde C}_k$ is a collection of $2k$ or $2k-1$ unit radius circles on this $\xi$-Riemann surface depending on $\alpha$. The first
value corresponds to $\alpha =1$ and the second to $\alpha =e^{-\frac{i\pi}{n}}$. The first circle starts at the point
$\xi=1$ on the sheet on which $\arg\xi=2k\pi$. The latter argument is just the one of $\xi_k=(-i\alpha z_k)^n$. In Fig.6
this is the $k$-th sheet from above with $a_{k+1}$ as the lower boundary of the cut between $0$ and $1$ on this sheet.
Each next
circle is an image of successive arcs of the contour $C_k$. The last circle corresponds to the unit radius arc between
the points $z_{-k+1}$ and $z_{-k}$ and it ends at unity on the $n-k+1$-th sheet.

As it follows from the Fig.6 the integration along ${\tilde C}_k$ on the $\xi$-Riemann surface can be deformed to go
along the cuts between the points $\xi=0$ and $\xi=1$ on the respective sheets of the surface. Then the integrations
along the $a_{k+1}$ and $a_{n-k+1}$ cuts survive only (the remaining ones mutually cancel) so that their
contributions to the last integral in \mref{A11} (call it  $I_k$) are followings

\be
I_k=-\frac{1}{i\alpha n}\int_{{\tilde C}_k}(\xi-1)^\fr\xi^\frac{1-n}{n}d\xi=\nn\\
\ll\{\ba{crc}
-\frac{1}{in}\int_0^1(1-\xi)^\fr\xi^\frac{1-n}{n}d\xi\ll(-e^{i\frac{n\pi+4k\pi}{2n}}+e^{i\frac{n\pi+4\pi(n-k+1)}{2n}}\r)&
for&\alpha=1\\
-\frac{1}{i\alpha n}\int_0^1(1-\xi)^\fr\xi^\frac{1-n}{n}d\xi\ll(-e^{i\frac{n\pi+4(k-1)\pi}{2n}}+e^{i\frac{n\pi+4\pi(n-k)}{2n}}\r)&
for&\alpha=e^{-\frac{i\pi}{n}}
\ea\r\}=\nn\\
\ll\{\ba{crc}
\frac{2i}{n}\sin\frac{2k\pi}{n}B\ll(\frac{3}{2},\frac{1}{n}\r)&for&\alpha=1,\;\;k=1,..., k<\frac{n}{2}\\
\frac{2i}{n}\sin\frac{(2k-1)\pi}{n}B\ll(\frac{3}{2},\frac{1}{n}\r)&
for&\alpha=e^{-\frac{i\pi}{n}},\;\;k=1,...,k\leq\frac{n}{2}
\ea\r\}
\label{A12}
\ee
where $B(x,y)$ is the betha function.

\section*{Appendix 2}

\hskip+2em Here we prove the {\bf Lemma} of sec.5, {\bf Theorems 2a-2b} of sec.5 and {\bf Lemma'} of sec.6.

{\bf Proof of Lemma}

To prove this lemma let us note that $\omega_\alpha(z)$ is holomorphic in $C_{cut}(\epsilon)$ vanishing there as $z^{-\fr n-2}$ for
$z\to\infty$. Therefore the integrals $I_k(z)=\int_{\gamma_k(z)}|\omega_\alpha(\xi)d\xi|,\;k=1,...n+2,$ are well defined
in the closure ${\bar C}_{cut}(\epsilon)$ and are bounded there. Hence $C_{\alpha;\epsilon}$ does exist and is finite in
${\bar C}_{cut}(\epsilon)$ and therefore can be taken the same for each $D_{k,\epsilon}$. The estimation \mref{14}
follows then directly from \mref{12} since
$\left|1 - e^{-2\sigma_k\lambda {\tilde W}_{n,\alpha}(y_i,y_{i+1})}\right|\leq 2$ in this formula for each
$i$ if all the integrations are performed on canonical paths $\gamma_k(z)$ what is allways possible by definition of
$D_{k,\epsilon}$.

{\bf Proof of Lemma'}

As we have mentioned in sec.6 the turning points $z_k(\lambda)$ of $W_n(z,\lambda)$ tend to corresponding roots $z_k$ of
$W_{n,\alpha}(z)$ in the limit $\lambda\to\infty$. Therefore taking $\lambda_0>0$ sufficiently large we can find all the
roots $z_k(\lambda)$ in the corresponding circles $|z-z_k|<\epsilon$ for $|\lambda|>\lambda_0$. Therefore as in the case
of $\omega_\alpha(z)$ also $\omega(z,\lambda)$
corresponding to the potential $W_n(z,\lambda)$ is holomorphic in the closure ${\bar C}_{cut}(\epsilon)$ and bounded there
since it vanishes there as $z^{-\fr n-2}$ for $z\to\infty$. Hence, as previously, the integrals
$I_k(z,\lambda)=\int_{\gamma_k(z)}|\omega_\alpha(\xi,\lambda)d\xi|,\;k=1,...n+2,$ are well defined in the closure
${\bar C}_{cut}(\epsilon)$ and are bounded there by $C_\epsilon(\lambda)$ defined by
\be
C_\epsilon(\lambda)=\liminf_{\gamma_k(z),z\in{\bar C}_{cut}(\epsilon),\;k=1,...,n+2}I_k(z,\lambda)
\label{A2a}
\ee

However $C_{\epsilon}(\lambda)$ is a continuous function of $\Re\lambda$ and $\Im\lambda$ for
$|\lambda|>\lambda_0$. Therefore since also $\lim_{\lambda\to\infty}C_{\epsilon}(\lambda)=C_{\alpha;\epsilon}$ in a closure of
${\bar C}_{cut}(\epsilon)$, $C_{\epsilon}(\lambda)$ is bounded by some
constant $C_{\epsilon}(\lambda_0)$ with the property that
$C_{\epsilon}(\lambda_0)\to C_{\alpha;\epsilon}$ for $\lambda_0\to\infty$. The latter means that for $\lambda_0$
sufficiently large we can choose a constant $C_{\epsilon}(\lambda_0)$ to be independent of $\lambda_0$ and to differ
from $C_{\alpha;\epsilon}$ by $\epsilon$, i.e. $C_{\epsilon}(\lambda_0)\leq C_{\epsilon}=
C_{\alpha;\epsilon}+\epsilon$.

Now the estimation \mref{161} follows directly from \mref{11} written for the potential $W_n(z,\lambda)+1$.

{\bf Proof of Theorem 2a}

According to {\bf Lemma}, for every $k$, $\psi_{\alpha,k}(z)$ can be represented in $D_{k,\epsilon}$ for
$|\lambda|>\lambda_0\gg C_{k,\epsilon}$ as
\be
\psi_{\alpha,k}(z)=
W_{n,\alpha}^{-\frac{1}{4}}(z)e^{\sigma_k\lambda\int_{z_k}^z\sqrt{W_{n,\alpha}(y)}dy}\ll(1+O(|\lambda|^{-\gamma})\r),
\;\;\;\;\gamma\geq 1
\label{A21}
\ee

In fact the above estimation of $\psi_{\alpha,k}(z)$ in $D_{k,\epsilon}$ can be extended to any finite order in
$\lambda^{-1}$. It will be convenient to include into this estimation all orders of $\lambda^{-1}$ i.e. to represent
$\psi_{\alpha,k}(z)$ by its full asymptotic form \mref{551}. Therefore we will use further the following
asymptotic representations for $\psi_{\alpha,k}(z)$:
\be
\psi_{\alpha,k}(z)\sim \psi_{\alpha,k}^{as}(z)=
W_{n,\alpha}^{-\frac{1}{4}}(z)e^{\sigma_k\lambda\int_{z_k}^z\sqrt{W_{n,\alpha}(y)}dy}\chi_k^{as}(z,\lambda)=\nn\\
W_{n,\alpha}^{-\frac{1}{4}}(z)e^{\sigma_k\lambda\int_{z_k}^z\sqrt{W_{n,\alpha}(y)}dy+\int_{\infty_k}^zZ_k(y,\lambda)dy}
\label{A211}
\ee

Dispite the fact that the asymptotic semiclassical series $Z_k(y,\lambda)$ is generally divergent we can use it in its
full form having however in mind that it can be abbreviated in any moment at some finite power of $\lambda^{-1}$ to
provide us with an estimation like \mref{A21} but valid then up to the order kept.

To perform detailed calculations we shall assume the $C_{cut}$-plane to be cut in the following way.

For the root $z_0$
and $z_{\frac{n}{2}}$ (if they are) the corresponding cuts coincide with the positive and negative parts of the imaginary
axis respectively. So there are in fact two SL's, mutually parallel to themselves on $C_{cut}$ and lying on different sites
of the cut.

For the remaining roots the corresponding cuts are parallel to the real axis running to the left for
the roots lying on the left from the imaginary axis and runnig to the right for the roots placed on the right from the
imaginary axis so that the pattern of cuts is completely symmetric with respect to the imaginary axis.

Arguments of
differences $z-z_k,\;k=0,\pm 1,...,n/2$, are always taken with respect to axis emerging from the roots $z_k$ parallelly
to the real axis and keeping its direction (see Fig.1a). For the right hand roots these arguments are taken from above the
cuts. The arguments are positive if they are taken clockwise and negative in the opposite cases.

With these conventions
a sign of $\Re\ll(\lambda\int_{z_k}^z\sqrt{W_{n,\alpha}(y)}dy\r),\;|\arg\lambda|\leq\frac{\pi}{n}$, is therefore
always positive for the sectors $S_0$, $S_{\pm\frac{n+3}{2}}$ and $S_{\pm\frac{n+2}{2}}$.
This sign is always negative for the sectors $S_{\frac{n+1}{2}}$ and $S_{\pm\frac{n}{2}}$. For the remaining
sectors $S_k,\;k=\pm 1,\pm 2,...,$ this sign is negative above the cuts crossing these sectors and positive below them,
i.e. it changes each time these cuts are crossed (see Fig.1a). Note that this switch of the signs does not occure on the
other cuts. In the same way behave all terms of the expansion $Z_k$ in \mref{A211}.

Also the factor $W_{n,\alpha}^{-\frac{1}{4}}(z)$ in \mref{A211} crossing each cut changes its phase
by $\mp i$. The upper sign has to be taken when the cut on the positive imaginary axis is crossed from its right or the
left hand cuts are crossed from above while the lower has to be taken for the negative imaginary axis cut and the r.h.
cuts crossed in the same directions (i.e. to the left and down respectively). The signs are changed to opposite if the
cuts are crossed in the opposite directions (see Fig.1a-1b).

Therefore we assume the form \mref{A211} of the solution $\psi_{\alpha,k}(z)$ to be valid in the negative parts of the
sectors $S_k,\;k=\pm 1,\pm 2,...,$ and in the sectors
$S_{\frac{n+1}{2}}$ and $S_{\pm\frac{n+2}{2}}$, so that $\sigma_k=+1$ there.

The form \mref{A211} is also assumed to be valid for the sectors $S_0$, $S_{\pm\frac{n+1}{2}}$ and
$S_{\pm\frac{n}{2}}$ but then $\sigma_k=-1$ for the corresponding solutions.

The asymptotic forms \mref{A211} of the FS's when the latter are continued analytically across the $C_{cut}$-plane along
canonical paths while keeping their validity change
therefore appropriately by acquiring additional phases by the factor $W_{n,\alpha}^{-\frac{1}{4}}(z)$ or by switching
their signs by the exponentials $\lambda\int_{z_k}^z\sqrt{W_{n,\alpha}(y)}dy+\int_{\infty_k}^zZ_k(y,\lambda)dy$ on the
cuts met.

We shall consider below in details the analytic continuation of the exact solutions $\psi_{\alpha,k}(z)$ and their
asymptotic expansions \mref{A211} as well along canonical paths. We will get first exact results of such continuations
to come over next to their asymptotic forms and to get finally their limit for $\lambda\to\infty$. We shall
consider them one by one beginning with $k=0,\frac{n+3}{2},\frac{n+2}{2},$
$1,2,...,\frac{n+1}{2}$, i.e. going down the corresponding SG's, but
neglecting detailed calculations for the solutions for which the corresponding results can be obtained be symmetry
arguments.

{\it $k=0$ case}

This case corresponds to an even-$n$ and $\alpha=e^{-\frac{i\pi}{n}}$ so that the corresponding SG with exceptional SL's
for this case is shown in Fig.1d-1e.

We shall continue $\psi_{\alpha,0}(z)$ to vicinities $V_{k,\epsilon}$ of its exceptional SL's (ESL) expressing it simply as
a linear combinations of FS's defined in sectors closest to the particular ESL considered at the moment. By symmetry
arguments we can do it only for the left ESL's of Fig.1d-1e.

Using first the exact forms \mref{10} of the solutions we can get coefficients of the coresponding linear combinations
also in their limit forms up to any order of $\lambda^{-1}$ by substituting the exact expressions by their asymptotics
\mref{A211}.

Therefore continuing $\psi_{\alpha,0}(z)$ to the vicinity of the ESL emerging from the root $z_k,\;k=1,...,\frac{n}{2}$
we get:
\be
\psi_{\alpha,0}(z)=ie^{-\lambda\int_{z_0}^{z_k}\sqrt{W_{n,\alpha}(y)}dy}
\frac{\chi_{0\to k+1}}{\chi_{k\to k+1}}\psi_{\alpha,k}(z)-\nn\\
ie^{-\lambda\int_{z_0}^{z_k}\sqrt{W_{n,\alpha}(y)}dy+\lambda\int_{z_k}^{z_{k+1}}\sqrt{W_{n,\alpha}(y)}dy}
\frac{\chi_{0\to k}}{\chi_{k\to k+1}}\psi_{\alpha,k+1}(z)=\nn\\
\frac{W_{n,\alpha}^{-\frac{1}{4}}(z)e^{-\lambda\int_{z_0}^{z_k}\sqrt{W_{n,\alpha}(y)}dy}}{\chi_{k\to k+1}}
\ll(e^{-\lambda\int_{z_k}^z\sqrt{W_{n,\alpha}(y)}dy}\chi_{0\to k+1}\chi_k(z,\lambda)\r.\nn\\
\ll.-ie^{\lambda\int_{z_k}^z\sqrt{W_{n,\alpha}(y)}dy}\chi_{0\to k}\chi_{k+1}(z,\lambda)\r)
\label{A22}
\ee
where $\chi_{i\to j}=\lim_{z\to\infty_j}\chi_i(z)=\lim_{z\to\infty_i}\chi_j(z)=\chi_{j\to i}$
(see, for example \cite{4}, ref.2).

From \mref{A22} we get the following exact condition for loci $\zeta_{k,m}^{(0)}$ of zeros of $\psi_{\alpha,0}(z)$ on the
considered ESL:
\be
\int_{z_k}^{\zeta_{k,m}^{(0)}}\sqrt{W_{n,\alpha}(y)}dy=+\ll(m-\frac{1}{4}\r)\frac{i\pi}{\lambda}+
\frac{1}{2\lambda}\ln\frac{\chi_{0\to k+1}\chi_k(\zeta_{k,m}^{(0)},\lambda)}
{\chi_{0\to k}\chi_{k+1}(\zeta_{k,m}^{(0)},\lambda)}
\label{A23}
\ee
where $m$ is positive integer and the choice of the "+"-sign is determined by the positiveness of the l.h.s. of
\mref{A23} on the ESL considered.

Now we would like to come over in \mref{A23} to the limit $\lambda\to\infty$. However if we want
to use the results of {\bf Lemma} we need $\zeta_{k,m}^{(0)}$ to be kept inside the domain $D_{k,\epsilon}$, i.e. their
distances to $z_k$ in the limit $\lambda\to\infty$ should be greater than $\epsilon$. It then follows from \mref{A23}
that for large $\lambda$ also $m$ should be large, i.e.
\be
\frac{m\pi}{|\lambda|}>
\limsup_{|\phi|\leq\pi}\ll|\int_{z_k}^{z_k+\epsilon e^{i\phi}}\sqrt{W_{n,\alpha}(y)}dy\r|\equiv I_k(\epsilon)
\label{A230}
\ee

This condition means that we have to take into account only those zeros $\zeta_{k,m}^{(0)}$ which are sufficiently far
from $z_k$. Therefore for those which can fall onto $z_k$ we have to take $m$ sufficiently large, i.e.
$m>m_0=|\lambda|I_k(\epsilon)/\pi$.

For the remaining ones, i.e. those which are to approach finite non-zero distances to
$z_k$ it is necessary to make $m$ growing linearly with
$\lambda$, i.e. $m=q[|\lambda|]+r$ with $q=1,2,...,\;r=0,\pm 1,\pm 2,...,$ and $[|\lambda|]$ being the
step function of $|\lambda|$, i.e. $|\lambda|=[|\lambda|]+\Lambda,\;0\leq\Lambda<1$.

Let us note further that the latter case will cover also the previous one when $q=0$ and $r>m_0$. In all the formulae below we
shall assume the cases of zeros falling down onto $z_k$ to be taken into account just in this way.

Having this in mind we can come in \mref{A23} to the limit $[|\lambda|]\to\infty$ to get for
$\zeta_{k,m}^{(0)}\sim\zeta_{k,qr}^{(0)}(\lambda)$:
\be
\int_{z_k}^{\zeta_{k,qr}^{(0)}(\lambda)}\sqrt{W_{n,\alpha}(y)}dy=+\ll(q[|\lambda|]+r-\frac{1}{4}\r)\frac{i\pi}{\lambda}+
\frac{1}{2\lambda}\int_{K_k(\zeta_{k,qr}^{(0)}(\Lambda))}Z_{\alpha,0}(y,\lambda)dy
\label{A231}
\ee
where $K_k(\zeta_{k,qr}^{(0)}(\lambda))$ is a contour shown in Fig.4a which starts and ends at the point
$\zeta_{k,qr}^{(0)}(\Lambda)$. This contour is not closed since it starts and ends at different sites of the cut
emerging from $z_k$.

In fact \mref{A231} is an implicit condition for the
asymptotic expansion $\zeta_{k,qr}^{(0)}(\lambda)$ of zeros $\zeta_{k,m}^{(0)}$
as $\lambda\to\infty$. Therefore we can look for
the following form of semiclassical expansion for $\zeta_{k,qr}^{(0)}(\lambda)$:
\be
\zeta_{k,qr}^{(0)}(\lambda)=\sum_{p\geq 0}\frac{1}{\lambda^p}\zeta_{k,qrp}^{(0)}(\Lambda)
\label{A232}
\ee
with $\zeta_{k,0r0}^{(0)}(\Lambda)\equiv z_k$.

Let us now note that the condition \mref{A231} can be written uniformly as:
\be
\int_{K_k(\zeta_{k,qr}^{(0)}(\lambda))}\ll(\fr\sqrt{W_{n,\alpha}(y)}-
\frac{1}{2\lambda}Z_{\alpha,0}(y,\lambda)\r)dy=+\ll(q[|\lambda|]+r-\frac{1}{4}\r)\frac{i\pi}{\lambda}
\label{A233}
\ee

We can now calculate the
asymptotic series \mref{A232} in the limit $[|\lambda|]\to\infty$ with fixed $\Lambda$ from the
following formula:
\be
\int_{K_k(\zeta_{k,qr0}^{(0)}(\Lambda))}\ll(\fr\sqrt{W_{n,\alpha}(y)}-
\frac{1}{2\lambda}Z_{\alpha,0}(y,\lambda)\r)dy+\nn\\
2\sum_{s\geq 1}\frac{1}{s!}\ll.\ll(\fr\sqrt{W_{n,\alpha}(y)}-
\frac{1}{2\lambda}Z_{\alpha,0}(y,\lambda)\r)^{(s)}\r|_{y=\zeta_{k,qr0}^{(0)}(\Lambda)}
\ll(\sum_{p\geq 1}\frac{1}{\lambda^p}\zeta_{k,qrp}^{(0)}(\Lambda)\r)^s=\nn\\
\ll(q[|\lambda|]+r-\frac{1}{4}\r)\frac{i\pi}{\lambda},\;\;\;q>0
\label{A234}
\ee
and with a similar formula for $q=0$ obtained from the last one where $\zeta_{k,qr0}^{(0)}(\Lambda)$ is substituted by
$\zeta_{k,0r1}^{(0)}(\Lambda)/\lambda$ and $p>1$.

The limit $[|\lambda|]\to\infty$ with fixed $\Lambda$ (and $\arg\lambda\equiv\beta,\;|\beta|<\pi/n$) considered above
is of course regular.

Therefore for $q>0$ the zero term $\zeta_{k,qr0}^{(0)}(\Lambda)$ of such an expansion is given by the equation:
\be
\int_{K_k(\zeta_{k,qr0}^{(0)}(\Lambda))}\fr\sqrt{W_{n,\alpha}(y)}dy=
\int_{z_k}^{\zeta_{k,qr0}^{(0)}(\Lambda)}\sqrt{W_{n,\alpha}(y)}dy=qi\pi e^{-i\beta}
\label{A235}
\ee
which shows that this term is independent of $\Lambda$.

The next term is given explicitely by:
\be
\zeta_{k,qr1}^{(0)}(\Lambda)=
(r-q\Lambda-\frac{1}{4})\frac{i\pi e^{-i\beta}}{\sqrt{W_{n,\alpha}(\zeta_{k,qr0}^{(0)}(\Lambda))}}
\label{A236}
\ee
and so on.

For $q=0$ we get correspondingly
\be
\int_{z_k}^{z_k+\zeta_{k,0r1}^{(0)}(\Lambda)/\lambda}\sqrt{W_{n,\alpha}(y)}dy=
(r-\frac{1}{4})\frac{i\pi}{\lambda},\;\;\;\;\;\;\;\;r>m_0=|\lambda|I_k(\epsilon)/\pi
\label{A237}
\ee
and
\be
\zeta_{k,0r2}^{(0)}(\Lambda)=\frac{1}{8}\frac{\int_{K_k(z_k+\zeta_{k,0r1}^{(0)}(\Lambda)/\lambda)}X_{\alpha,1}(y)dy}
{\sqrt{W_{n,\alpha}(z_k+\zeta_{k,0r1}^{(0)}(\Lambda)/\lambda)}}
\label{A238}
\ee

{\it $k=\frac{n+3}{2}$ case}

This case corresponds to $n$ odd and to the pattern of ESL's shown in Fig.1a. We have to continue
$\psi_{\alpha,\frac{n+3}{2}}(z)$ to vicinities of ESL's emerging from the root $z_0$ and from the roots
$z_k,\;k=\pm 1,\pm 2,...,\pm\frac{n-1}{2}$. In the first case we have to express $\psi_{\alpha,\frac{n+3}{2}}(z)$ linearly by
the solutions $\psi_{\alpha,-\frac{n+3}{2}}(z)$ and $\psi_{\alpha,-1}(z)$. For the remaining ESL's
$\psi_{\alpha,\frac{n+3}{2}}(z)$ is continued in the same way as in the previous case. We have therefore:
\be
\psi_{\alpha,\frac{n+3}{2}}(z)=\nn\\\frac{\chi_{\frac{n+3}{2}\to -1}}{\chi_{-1\to -\frac{n+3}{2}}}
\psi_{\alpha,-\frac{n+3}{2}}(z)+
ie^{\lambda\int_{z_0}^{z_{-1}}\sqrt{W_{n,\alpha}(y)}dy}
\frac{\chi_{\frac{n+3}{2}\to -\frac{n+3}{2}}}{\chi_{-1\to -\frac{n+3}{2}}}\psi_{\alpha,-1}(z)
\label{A24}
\ee
for the first ESL and
\be
\psi_{\alpha,\frac{n+3}{2}}(z)=\nn\\
ie^{-\lambda\int_{z_0}^{z_k}\sqrt{W_{n,\alpha}(y)}dy}\frac{\chi_{\frac{n+3}{2}\to k+1}}{\chi_{k\to k+1}}
\psi_{\alpha,k}(z)-\nn\\
ie^{-\lambda\int_{z_0}^{z_k}\sqrt{W_{n,\alpha}(y)}dy+\lambda\int_{z_k}^{z_{k+1}}\sqrt{W_{n,\alpha}(y)}dy}
\frac{\chi_{\frac{n+3}{2}\to k}}{\chi_{k\to k+1}}\psi_{\alpha,k+1}(z)
\label{A25}
\ee
for the ESL's emerging from the roots $z_k,\;k=1,2,...,\frac{n-1}{2}$ while for the ES's emerging from the roots with the
opposite sign of $k$ we get:
\be
\psi_{\alpha,\frac{n+3}{2}}(z)=\nn\\
e^{-\lambda\int_{z_0}^{z_{-k}}\sqrt{W_{n,\alpha}(y)}dy}\frac{\chi_{\frac{n+3}{2}\to -k-1}}{\chi_{-k\to -k-1}}
\psi_{\alpha,-k}(z)+\nn\\
ie^{-\lambda\int_{z_0}^{z_{-k}}\sqrt{W_{n,\alpha}(y)}dy+\lambda\int_{z_{-k}}^{z_{-k-1}}\sqrt{W_{n,\alpha}(y)}dy}
\frac{\chi_{\frac{n+3}{2}\to -k}}{\chi_{-k\to -k-1}}\psi_{\alpha,-k-1}(z)
\label{A26}
\ee

Arguing in exactly the same way as in the previous case from \mref{A24} and \mref{A26} for zeros of
$\psi_{\alpha,\frac{n+3}{2}}(z)$ on the ESL's emerging from the roots $z_{\pm k},\;k=0,1,2,...,\frac{n-1}{2}$ we get
the corresponding conditions:
\be
\int_{K_{\pm k}(\zeta_{\pm k,qr}^{(\frac{n+3}{2})}(\Lambda))}\ll(\fr\sqrt{W_{n,\alpha}(y)}-
\frac{1}{2\lambda}Z_{\alpha,\frac{n+3}{2}}(y,\lambda)\r)dy=\pm\ll(q[|\lambda|]+r-\frac{1}{4}\r)\frac{i\pi}{\lambda}
\label{A27}
\ee

{\it $k=\frac{n+2}{2}$ case}

This case corresponds to the even $n$ and $\alpha=1$, see Fig.1b. We have to continue $\psi_{\alpha,\frac{n+2}{2}}(z)$
to vicinities of its the ESL's emerging from the roots $z_{\pm k},\;k=0,1,...,\frac{n-2}{2}$ and from $z_\frac{n}{2}$.

However,
the corresponding formulae for the positions of zeros on the ESL's emerging from all the roots mentioned above except the
last one coincide for the obvious reasons exactly with the respective formulae \mref{A27} while for the ESL's emerging
from the last root we have
\be
\psi_{\alpha,\frac{n+2}{2}}(z)=\nn\\
ie^{-\lambda\int_{z_0}^{z_{\frac{n}{2}}}\sqrt{W_{n,\alpha}(y)}dy}\ll(\chi_{\frac{n+3}{2}\to -\frac{n}{2}}
\psi_{\alpha,\frac{n}{2}}(z)-
\chi_{\frac{n+3}{2}\to\frac{n}{2}}\psi_{\alpha,-\frac{n}{2}}(z)\r)
\label{A28}
\ee
where $z$ is assumed to lie on the l.h.s. of the cut emerging from $z_{\frac{n}{2}}$.

Therefore for zeros
$\zeta_{\frac{n}{2},qr}^{(\frac{n+2}{2})}$ of $\psi_{\alpha,\frac{n+2}{2}}(z)$ lying in vicinity of the ESL emerging from
$z_{\frac{n}{2}}$ we get:
\be
\int_{K_{\frac{n}{2}}(\zeta_{\frac{n}{2},qr}^{(\frac{n+2}{2})}(\Lambda))}\ll(\fr\sqrt{W_{n,\alpha}(y)}-
\frac{1}{2\lambda}Z_{\alpha,\frac{n+2}{2}}(y,\lambda)\r)dy=\pm\ll(q[|\lambda|]+r-\frac{1}{4}\r)\frac{i\pi}{\lambda}
\label{A29}
\ee
where $\pm$ corresponds to the positions of $\zeta_{\frac{n}{2},qr}^{(\frac{n+2}{2})}$ on the left or right from the cut
respectively. Of course we always get the same values for $\zeta_{\frac{n}{2},qr}^{(\frac{n+2}{2})}$ irrespectively of the
site chosen.

{\it $k=1,2,...,\frac{n}{2}$ cases}

We shall consider all the enumerated cases jointly since the corresponding roots are all on the left from the imaginary axis
independently of the parity of $n$.

We have to distinguish two cases corresponding to $\arg\lambda=\beta$ is equal to $0$ or differs from it. In the second
case (the corresponding SG is shown in Fig.2)
the procedure of analytical continuations of $\psi_{\alpha,k}(z)$ along canonical paths to vicinities of all its the
ESL's is completely analogous to the previous cases since all the sectors having these ESL's as their boundaries are
available from the sector $S_k$ along canonical paths. The rules of making these continuations are now clear and
their results are gathered by the formulae of the type \mref{A233}, \mref{A27} or \mref{A29} so we can focuse our attention on the
real case of $\lambda$.

In the case $\beta=0$ each inner SL emerging from the root $z_k,\;k=1,2,...,\frac{n}{2}$ is exceptional for the
corresponding solution $\psi_{\alpha,k}(z)$, see Fig.1a-1e, together with the two ESL's which close the sector $S_{-k}$
which is not available directly on any canonical paths from the sector $S_k$.

On the other hand positions of other ESL's
corresponding to $\psi_{\alpha,k}(z)$ in the considered case is analogous to their positions for $\beta\neq 0$ and their
vicinities can be approached by continuations of $\psi_{\alpha,k}(z)$ along canonical paths, i.e. positions of zeros of
the considered solution on these ELS's can be established exactly in the same way as for the cases investigated so far.
Therefore for positions of these zeros on their ESL's we can invoke again the respective formulae \mref{A233}, \mref{A27}
and \mref{A29}.

Consider now positions of possible zeros of $\psi_{\alpha,k}(z)$ on the inner line emerging from the root $z_k$ (it ends
at $z_{-k}$) and on the remaining two lines emerging from $z_{-k}$ and runnig to infinity. For the inner line we can use
any pair of linear independent FS's for both of which this line is not exceptional. These can be for example the
solutions $\psi_{\alpha,-k+1}(z)$ and $\psi_{\alpha,-k-1}(z)$. They communicate canonically with the solution
$\psi_{\alpha,k}(z)$ so we get:
\be
\psi_{\alpha,k}(z)=
-e^{-\lambda\int_{z_k}^{z_{-k+1}}\sqrt{W_{n,\alpha}(y)}dy}
\frac{\chi_{k\to -k-1}}{\chi_{-k+1\to -k-1}}\psi_{\alpha,-k+1}(z)+\nn\\
e^{\lambda\int_{z_k}^{z_{-k-1}}\sqrt{W_{n,\alpha}(y)}dy}\frac{\chi_{k\to -k+1}}{\chi_{-k+1\to -k-1}}\psi_{\alpha,-k-1}(z)
\label{A30}
\ee

From \mref{A30} we get for the positions of zeros of $\psi_{\alpha,k}(z)$ on its inner ESL $(z_k,z_{-k})$:
\be
\int_{K_{k}(\zeta_{k,qr}^{(k)}(\Lambda))}\ll(\fr\sqrt{W_{n,\alpha}(y)}-
\frac{1}{2\lambda}Z_{\alpha,k}(y,\lambda)\r)dy=-\ll(m-\frac{1}{4}\r)\frac{i\pi}{\lambda}
\label{A31}
\ee

It follows from \mref{A235} that $q$ is limited now by the integral
$I_k\equiv\int_{z_k}^{z_{-k}}\sqrt{W_{n,\alpha}(y)}dy$, i.e. $q\leq-I_k/(i\pi)$.

In order to look for possible zeros of $\psi_{\alpha,k}(z)$ on its ESL's emerging from $z_{-k}$ we have to continue it
further to vicinities of these lines using the solution $\psi_{\alpha,-k}(z)$ which also comunicates canonically with
both the solutions $\psi_{\alpha,-k+1}(z)$ and $\psi_{\alpha,-k-1}(z)$. Choosing one of them we can express it as linear
combinations of the others to get:
\be
\psi_{\alpha,-k+1}(z)=\nn\\
e^{\lambda\int_{z_{-k}}^{z_{-k+1}}\sqrt{W_{n,\alpha}(y)}dy}
\chi_{-k+1\to -k-1}\psi_{\alpha,-k}(z)-\nn\\
e^{\lambda\int_{z_{-k}}^{z_{-k+1}}\sqrt{W_{n,\alpha}(y)}dy+\lambda\int_{z_{-k}}^{z_{-k-1}}\sqrt{W_{n,\alpha}(y)}dy}
\psi_{\alpha,-k-1}(z)
\label{A32}
\ee
and
\be
\psi_{\alpha,-k-1}(z)=\nn\\
e^{-\lambda\int_{z_{-k}}^{z_{-k-1}}\sqrt{W_{n,\alpha}(y)}dy}
\chi_{-k-1\to -k+1}\psi_{\alpha,-k}(z)-\nn\\
e^{-\lambda\int_{z_{-k}}^{z_{-k+1}}\sqrt{W_{n,\alpha}(y)}dy-\lambda\int_{z_{-k}}^{z_{-k-1}}\sqrt{W_{n,\alpha}(y)}dy}
\psi_{\alpha,-k+1}(z)
\label{A33}
\ee
respectively.

Now substituting subsequently both the above formulae to \mref{A30} we get the formulae realizing continuations of
$\psi_{\alpha,k}(z)$ close to the respective ELS's, i.e. we get in this way:
\be
\psi_{\alpha,k}(z)=-e^{-\lambda\int_{z_k}^{z_{-k}}\sqrt{W_{n,\alpha}(y)}dy}\frac{\chi_{k\to -k-1}}{\chi_{-k+1\to -k-1}}
\ll(\chi_{-k+1\to -k-1}\psi_{\alpha,-k}(z)-\r.\nn\\
\ll.e^{\lambda\int_{z_{-k}}^{z_{-k-1}}\sqrt{W_{n,\alpha}(y)}dy}
\ll(1+e^{2\lambda\int_{z_k}^{z_{-k}}\sqrt{W_{n,\alpha}(y)}dy}\frac{\chi_{k\to -k+1}}
{\chi_{k\to -k-1}}\r)\psi_{\alpha,-k-1}(z)\r)
\label{A34}
\ee
and
\be
\psi_{\alpha,k}(z)=+e^{\lambda\int_{z_k}^{z_{-k}}\sqrt{W_{n,\alpha}(y)}dy}
\frac{\chi_{k\to -k+1}}{\chi_{-k+1\to -k-1}}\ll(\chi_{-k+1\to -k-1}\psi_{\alpha,-k}(z)-\r.\nn\\
\ll.e^{-\lambda\int_{z_{-k}}^{z_{-k+1}}\sqrt{W_{n,\alpha}(y)}dy}
\ll(1+
e^{-2\lambda\int_{z_k}^{z_{-k}}\sqrt{W_{n,\alpha}(y)}dy}\frac{\chi_{k\to -k-1}}
{\chi_{k\to -k+1}}\r)\psi_{\alpha,-k+1}(z)\r)
\label{A35}
\ee

The first of these formulae is suitable for analysing zeros of $\psi_{\alpha,k}(z)$ in a vicinity of the SL being the
lower boundary of the sector $S_{-k}$ while the second in vicinity of the SL being its upper boundary.

The first \mref{A34} of these formulae provide us with the following condition for zeros $\zeta_{-k,m}^{(k)-}$ of
$\psi_{\alpha,k}(z)$ in the limit $\lambda\to\infty$:
\be
\int_{z_{-k}}^{\zeta_{-k,m}^{(k)-}}\sqrt{W_{n,\alpha}(y)}dy=-\ll(m-\frac{1}{4}\r)\frac{i\pi}{\lambda}+
\frac{1}{4\lambda}\ll(\lambda\oint_{K_k}\sqrt{W_{n,\alpha}(y)}dy+\oint_{K_k}Z_{\alpha,k}dy\r)-\nn\\
\frac{1}{2\lambda}\ln2\cos\fr\Im\ll(\lambda\oint_{K_k}\sqrt{W_{n,\alpha}(y)}dy+\oint_{K_k}Z_{\alpha,k}dy\r)-
\frac{1}{2\lambda}\int_{K_{-k}(\zeta_{-k,m}^{(k)-})}Z_{\alpha,k}dy
\label{A36}
\ee
where $K_k$ and $K_{-k}(\zeta_{-k,m}^{(k)-})$ are contours shown in Fig.

Putting now $\lambda_{r_k}\oint_{K_k}\sqrt{W_{n,\alpha}(y)}dy=2(r_k+R)i\pi,\;|R|<\fr$ with $r_k$ a non-negative
integer, $\lambda_{r_k}=[\lambda_{r_k}]+\Lambda_{r_k}(R),\;0\leq\Lambda_{r_k}<1$ and $m=q[\lambda_{r_k}]+r$ we get from
\mref{A36}:
\be
\int_{z_{-k}}^{\zeta_{-k,qr}^{(k)-}}\sqrt{W_{n,\alpha}(y)}dy=-\ll(q[\lambda_{r_k}]+r-\frac{1}{4}+\frac{R}{2}\r)
\frac{i\pi}{\lambda_{r_k}}+
\frac{1}{4\lambda_{r_k}}\oint_{K_k}Z_{\alpha,k}dy-\nn\\
\frac{1}{2\lambda_{r_k}}\ln2\cos\ll(R\pi+\fr\Im\oint_{K_k}Z_{\alpha,k}dy\r)-\frac{1}{2\lambda_{r_k}}\int_{K_{-k}(\zeta_{-k,qr}^{(k)-})}Z_{\alpha,k}dy
\label{A361}
\ee
where the limit taken is regular with fixed $R$.

Therefore for the first two terms of the semiclassical expansion of $\zeta_{-k,qr}^{(k)-}$ we get:
\be
\int_{z_{-k}}^{\zeta_{-k,qr0}^{(k)-}}\sqrt{W_{n,\alpha}(y)}dy=-qi\pi\nn\\
\zeta_{-k,qr1}^{(k)-}(R)=-\ll(r-q\Lambda_{r_k}(R)-\frac{1}{4}+\frac{R}{2}
+\fr\ln2\cos(R\pi)\r)\frac{i\pi}{\sqrt{W_{n,\alpha}(\zeta_{-k,qr0}^{(k)-})}}
\label{A362}
\ee

In exactly the same way we get the condition for zeros $\zeta_{-k,m}^{(k)+}$ distributed along the upper SL emerging
from the turning point $z_{-k}$ in the limit $\lambda\to\infty$:
\be
\int_{z_{-k}}^{\zeta_{-k,m}^{(k)+}}\sqrt{W_{n,\alpha}(y)}dy=-\ll(m-\frac{1}{4}\r)\frac{i\pi}{\lambda}+
\frac{1}{4\lambda}\ll(\lambda\oint_{K_k}\sqrt{W_{n,\alpha}(y)}dy+\oint_{K_k}Z_{\alpha,k}dy\r)+\nn\\
\frac{1}{2\lambda}\ln2\cos\fr\Im\ll(\lambda\oint_{K_k}\sqrt{W_{n,\alpha}(y)}dy+\oint_{K_k}Z_{\alpha,k}dy\r)+
\frac{1}{2\lambda}\int_{K_{-k}(\zeta_{-k,m}^{(k)+})}Z_{\alpha,k}dy
\label{A37}
\ee
and with the same meaning of notations used we get:
\be
\int_{z_{-k}}^{\zeta_{-k,qr}^{(k)+}}\sqrt{W_{n,\alpha}(y)}dy=-\ll(q[\lambda_{r_k}]+r-\frac{1}{4}+\frac{R}{2}\r)
\frac{i\pi}{\lambda_{r_k}}+
\frac{1}{4\lambda_{r_k}}\oint_{K_k}Z_{\alpha,k}dy+\nn\\
\frac{1}{2\lambda_{r_k}}\ln2\cos\ll(R\pi+\fr\Im\oint_{K_k}Z_{\alpha,k}dy\r)+
\frac{1}{2\lambda_{r_k}}\int_{K_{-k}(\zeta_{-k,qr}^{(k)+})}Z_{\alpha,k}dy
\label{A371}
\ee
with the following first coefficients of the corresponding semiclassical expansion of $\zeta_{-k,qr}^{(k)+}$:
\be
\int_{z_{-k}}^{\zeta_{-k,qr0}^{(k)+}}\sqrt{W_{n,\alpha}(y)}dy=-qi\pi\nn\\
\zeta_{-k,qr1}^{(k)+}(R)=-\ll(r-q\Lambda_{r_k}(R)-\frac{1}{4}+\frac{R}{2}
-\fr\ln2\cos(R\pi)\r)\frac{i\pi}{\sqrt{W_{n,\alpha}(\zeta_{-k,qr0}^{(k)+})}}
\label{A372}
\ee

{\it $k=\frac{n+1}{2}$ case}

This is the last distinguished case which has to be treated. It corresponds to an odd $n$ of Fig.1a and to the solution
$\psi_{\alpha,\frac{n+1}{2}}(z)$. We should continue the latter solution to its ESL's emerging from the roots
$z_k,\;k=1,...,\frac{n-1}{2}$ and from $z_0$. Continuing to any ESL of the first group we get:
\be
\psi_{\alpha,\frac{n+1}{2}}(z)=
e^{\lambda\int_{z_\frac{n+1}{2}}^{z_k}\sqrt{W_{n,\alpha}(y)}dy}\chi_{\frac{n+1}{2}\to k-1}\psi_{\alpha,k}(z)-\nn\\
e^{\lambda\int_{z_\frac{n+1}{2}}^{z_k}\sqrt{W_{n,\alpha}(y)}dy-\lambda\int_{z_k}^{z_{k-1}}\sqrt{W_{n,\alpha}(y)}dy}
\chi_{\frac{n+1}{2}\to k}\psi_{\alpha,k-1}(z)=
W_{n,\alpha}^{-\frac{1}{4}}(z)e^{\lambda\int_{z_\frac{n+1}{2}}^{z_k}\sqrt{W_{n,\alpha}(y)}dy}\times\nn\\
\ll(e^{\lambda\int_{z_k}^z\sqrt{W_{n,\alpha}(y)}dy}\chi_{\frac{n+1}{2}\to k-1}\chi_k(z)+
ie^{-\lambda\int_{z_k}^z\sqrt{W_{n,\alpha}(y)}dy}\chi_{\frac{n+1}{2}\to k}\chi_{k-1}(z)\r)
\label{A38}
\ee
and hence for the respective distributions of zeros $\zeta_{k,qr}^{(\frac{n+1}{2})}(\Lambda)$:
\be
\int_{K_{k}(\zeta_{k,qr}^{(\frac{n+1}{2})}(\Lambda))}\ll(\fr\sqrt{W_{n,\alpha}(y)}-
\frac{1}{2\lambda}Z_{\alpha,\frac{n+1}{2}}(y,\lambda)\r)dy=\ll(q[|\lambda|]+r-\frac{1}{4}\r)\frac{i\pi}{\lambda}
\label{A39}
\ee

For the ESL emerging from $z_0$ we have:
\be
\psi_{\alpha,\frac{n+1}{2}}(z)=\nn\\
ie^{\lambda\int_{z_\frac{n+1}{2}}^{z_0}\sqrt{W_{n,\alpha}(y)}dy}\ll(\chi_{\frac{n+1}{2}\to\frac{n+3}{2}}\psi_{\alpha,\frac{n+3}{2}}(z)-
\chi_{\frac{n+1}{2}\to-\frac{n+3}{2}}\psi_{\alpha,-\frac{n+3}{2}}(z)\r)=\nn\\
iW_{n,\alpha}^{-\frac{1}{4}}(z)e^{\lambda\int_{z_\frac{n+1}{2}}^{z_0}\sqrt{W_{n,\alpha}(y)}dy}
\ll(e^{-\lambda\int_{z_0}^z\sqrt{W_{n,\alpha}(y)}dy}\chi_{\frac{n+1}{2}\to\frac{n+3}{2}}\chi_{\frac{n+3}{2}}(z)+\r.\nn\\
\ll.ie^{\lambda\int_{z_0}^z\sqrt{W_{n,\alpha}(y)}dy}\chi_{\frac{n+1}{2}\to-\frac{n+3}{2}}\chi_{-\frac{n+3}{2}}(z)\r)
\label{A40}
\ee
and hence:
\be
\int_{K_{0}(\zeta_{0,qr}^{(\frac{n+1}{2})}(\Lambda))}\ll(\fr\sqrt{W_{n,\alpha}(y)}-
\frac{1}{2\lambda}Z_{\alpha,\frac{n+1}{2}}(y,\lambda)\r)dy=\ll(q[|\lambda|]+r-\frac{1}{4}\r)\frac{i\pi}{\lambda}
\label{A41}
\ee
for the respective distributions of zeros on the considered ESL.

\section*{Appendix 3}

\hskip+2em We show here that for the potential $W_n(z,\lambda)$ in the limit $\lambda\to\infty$ if there is a unique
inner SL of the SG for this potential between the turning points $z_{k_0}(\lambda)$ and $z_{-k_0}(\lambda)$ then
$\arg\lambda\neq 0$
and in general case $\arg\lambda\sim|\lambda|^{-\frac{2}{n+2}}$. To see this consider in the limit $\lambda\to\infty$ the
integral $\lambda\int_{z_{k_0}(\lambda)}^{z_{-k_0}(\lambda)}\sqrt{W_n(y,\lambda)dy}$. With the accuracy
$O\ll(|\lambda|^{-\frac{3}{n+2}}\r)$ we have:
\be
\lambda\int_{z_{k_0}(\lambda)}^{z_{-k_0}(\lambda)}\sqrt{W_n(y,\lambda)}dy=
\lambda\int_{z_{k_0}}^{z_{-k_0}}\sqrt{W_{\alpha,n}(y)}dy-\nn\\
\frac{1}{2n}\lambda b_{n-1}'(-i\alpha)^{n-2}\int_{z_{k_0}}^{z_{-k_0}}\frac{y^{n-1}dy}{\sqrt{W_{\alpha,n}(y)}}
\frac{1}{\lambda^\frac{2}{n+2}}
\label{A42}
\ee
where we have used a relation $z_k(\lambda)=z_k-i\frac{1}{n\alpha}b_{n-1}'\lambda^{-\frac{2}{n+2}}+
O(\lambda^{-\frac{4}{n+2}})$ valid for any turning point in this limit, with $z_k$ being a turning point of
$W_{\alpha,n}(z)$.

According to our assumption we have further ($\lambda=|\lambda|e^{i\beta}$):
\be
\Re\ll(i\beta\int_{z_{k_0}(\lambda)}^{z_{-k_0}(\lambda)}\sqrt{W_n(y,\lambda)}dy\r)=-\frac{1}{2n}
\Re\ll(b_{n-1}'(-i\alpha)^{n-2}\int_{z_{k_0}}^{z_{-k_0}}\frac{y^{n-1}}{\sqrt{W_{\alpha,n}(y)}dy}\r)
\frac{1}{|\lambda|^\frac{2}{n+2}}
\label{A43}
\ee
and hence finally:
\be
\beta=-\frac{1}{2niI_{k_0}}
\Re\ll(b_{n-1}'(-i\alpha)^{n-2}\int_{z_{k_0}}^{z_{-k_0}}\frac{y^{n-1}}{\sqrt{W_{\alpha,n}(y)}dy}\r)
\frac{1}{|\lambda|^\frac{2}{n+2}}
\label{A44}
\ee
where $I_{k_0}=\int_{z_{k_0}}^{z_{-k_0}}\sqrt{W_{n,\alpha}(y)}dy$.

\end{document}